\documentclass[12pt,dvips]{article}
\usepackage{rotating}
\usepackage{axodraw}
\usepackage{epsfig}
\usepackage{psfig}
\usepackage{color}
\usepackage{cite}
\definecolor{GreenYellow}   {cmyk}{0.15,0,0.69,0}
\definecolor{Yellow}        {cmyk}{0,0,1,0}
\definecolor{Goldenrod}     {cmyk}{0,0.10,0.84,0}
\definecolor{Dandelion}     {cmyk}{0,0.29,0.84,0}
\definecolor{Apricot}       {cmyk}{0,0.32,0.52,0}
\definecolor{Peach}         {cmyk}{0,0.50,0.70,0}
\definecolor{Melon}         {cmyk}{0,0.46,0.50,0}
\definecolor{YellowOrange}  {cmyk}{0,0.42,1,0}
\definecolor{Orange}        {cmyk}{0,0.61,0.87,0}
\definecolor{BurntOrange}   {cmyk}{0,0.51,1,0}
\definecolor{Bittersweet}   {cmyk}{0,0.75,1,0.24}
\definecolor{RedOrange}     {cmyk}{0,0.77,0.87,0}
\definecolor{Mahogany}      {cmyk}{0,0.85,0.87,0.35}
\definecolor{Maroon}        {cmyk}{0,0.87,0.68,0.32}
\definecolor{BrickRed}      {cmyk}{0,0.89,0.94,0.28}
\definecolor{Red}           {cmyk}{0,1,1,0}
\definecolor{OrangeRed}     {cmyk}{0,1,0.50,0}
\definecolor{RubineRed}     {cmyk}{0,1,0.13,0}
\definecolor{WildStrawberry}{cmyk}{0,0.96,0.39,0}
\definecolor{Salmon}        {cmyk}{0,0.53,0.38,0}
\definecolor{CarnationPink} {cmyk}{0,0.63,0,0}
\definecolor{Magenta}       {cmyk}{0,1,0,0}
\definecolor{VioletRed}     {cmyk}{0,0.81,0,0}
\definecolor{Rhodamine}     {cmyk}{0,0.82,0,0}
\definecolor{Mulberry}      {cmyk}{0.34,0.90,0,0.02}
\definecolor{RedViolet}     {cmyk}{0.07,0.90,0,0.34}
\definecolor{Fuchsia}       {cmyk}{0.47,0.91,0,0.08}
\definecolor{Lavender}      {cmyk}{0,0.48,0,0}
\definecolor{Thistle}       {cmyk}{0.12,0.59,0,0}
\definecolor{Orchid}        {cmyk}{0.32,0.64,0,0}
\definecolor{DarkOrchid}    {cmyk}{0.40,0.80,0.20,0}
\definecolor{Purple}        {cmyk}{0.45,0.86,0,0}
\definecolor{Plum}          {cmyk}{0.50,1,0,0}
\definecolor{Violet}        {cmyk}{0.79,0.88,0,0}
\definecolor{RoyalPurple}   {cmyk}{0.75,0.90,0,0}
\definecolor{BlueViolet}    {cmyk}{0.86,0.91,0,0.04}
\definecolor{Periwinkle}    {cmyk}{0.57,0.55,0,0}
\definecolor{CadetBlue}     {cmyk}{0.62,0.57,0.23,0}
\definecolor{CornflowerBlue}{cmyk}{0.65,0.13,0,0}
\definecolor{MidnightBlue}  {cmyk}{0.98,0.13,0,0.43}
\definecolor{NavyBlue}      {cmyk}{0.94,0.54,0,0}
\definecolor{RoyalBlue}     {cmyk}{1,0.50,0,0}
\definecolor{Blue}          {cmyk}{1,1,0,0}
\definecolor{Cerulean}      {cmyk}{0.94,0.11,0,0}
\definecolor{Cyan}          {cmyk}{1,0,0,0}
\definecolor{ProcessBlue}   {cmyk}{0.96,0,0,0}
\definecolor{SkyBlue}       {cmyk}{0.62,0,0.12,0}
\definecolor{Turquoise}     {cmyk}{0.85,0,0.20,0}
\definecolor{TealBlue}      {cmyk}{0.86,0,0.34,0.02}
\definecolor{Aquamarine}    {cmyk}{0.82,0,0.30,0}
\definecolor{BlueGreen}     {cmyk}{0.85,0,0.33,0}
\definecolor{Emerald}       {cmyk}{1,0,0.50,0}
\definecolor{JungleGreen}   {cmyk}{0.99,0,0.52,0}
\definecolor{SeaGreen}      {cmyk}{0.69,0,0.50,0}
\definecolor{Green}         {cmyk}{1,0,1,0}
\definecolor{ForestGreen}   {cmyk}{0.91,0,0.88,0.12}
\definecolor{PineGreen}     {cmyk}{0.92,0,0.59,0.25}
\definecolor{LimeGreen}     {cmyk}{0.50,0,1,0}
\definecolor{YellowGreen}   {cmyk}{0.44,0,0.74,0}
\definecolor{SpringGreen}   {cmyk}{0.26,0,0.76,0}
\definecolor{OliveGreen}    {cmyk}{0.64,0,0.95,0.40}
\definecolor{RawSienna}     {cmyk}{0,0.72,1,0.45}
\definecolor{Sepia}         {cmyk}{0,0.83,1,0.70}
\definecolor{Brown}         {cmyk}{0,0.81,1,0.60}
\definecolor{Tan}           {cmyk}{0.14,0.42,0.56,0}
\definecolor{Gray}          {cmyk}{0,0,0,0.50}
\definecolor{Black}         {cmyk}{0,0,0,1}
\definecolor{White}         {cmyk}{0,0,0,0}

\newcommand{\wt}{\widetilde}
\newcommand{\imag}{\Im {\rm m}}
\newcommand{\real}{\Re {\rm e}}

\newcommand{\lsim}{\raisebox{-0.13cm}{~\shortstack{$<$ \\[-0.07cm] $\sim$}}~}

\begin{document}

\def\thefootnote{\fnsymbol{footnote}}

\begin{flushright}
ANL-HEP-PR-03-060,
CERN-TH/2003-156,
EFI-03-38,\\
FERMILAB-Pub-03/215-T,
KIAS-P03055,
MC-TH-2003-06,
TUM-HEP-517/03\\
hep-ph/0307377,
July 2003
\end{flushright}

\begin{center}
{\bf {\LARGE
{\color{Red}CP}{\color{Blue}super}{\color{OliveGreen}H}\hspace{1mm}:}
}\\[3.mm]
{\bf {\Large a Computational Tool for Higgs Phenomenology }}\\[2.5mm]
{\bf {\Large in the Minimal Supersymmetric Standard Model}}\\[2.5mm]
{\bf {\Large with Explicit CP Violation } }
\end{center}

\medskip

\begin{center}{\large
J.~S.~Lee$^a$,
A.~Pilaftsis$^a$,
M.~Carena$^b$,
S.~Y.~Choi$^{c}$,
}\\[0.2cm]
{\large
M.~Drees$^d$,
J.~Ellis$^e$
and C.~E.~M.~Wagner$^{f,g}$}
\end{center}

\begin{center}
{\em $^a$Department of Physics and Astronomy, University of Manchester}\\
{\em Manchester M13 9PL, United Kingdom}\\[0.2cm]
{\em $^b$Fermilab, P.O. Box 500, Batavia IL 60510, U.S.A.}\\[0.2cm]
{\em $^c$Physics Department, Chonbuk National University, Chonju
561--756, Korea}\\[0.2cm]
{\em $^d$Physik Department, Technische Universit\"at M\"unchen,
        D-85748 Garching, Germany}\\[0.2cm]
{\em $^e$Theory Division, CERN, CH-1211 Geneva 23, Switzerland}\\[0.2cm]
{\em $^f$HEP Division, Argonne National Laboratory,
9700 Cass Ave., Argonne, IL 60439, USA}\\[0.2cm]
{\em $^g$Enrico Fermi Institute, Univ. of Chicago, 5640
Ellis Ave., Chicago, IL 60637, USA}
\end{center}

\bigskip 

\centerline{\bf ABSTRACT}
\medskip\noindent  We provide  a detailed  description of  the Fortran
code  {\tt CPsuperH},  a newly--developed  computational  package that
calculates  the mass  spectrum and  decay  widths of  the neutral  and
charged Higgs bosons in the Minimal Supersymmetric Standard Model with
explicit   CP   violation.   The    program   is   based   on   recent
renormalization-group-improved diagrammatic  calculations that include
dominant   higher--order   logarithmic   and  threshold   corrections,
$b$-quark   Yukawa-coupling   resummation   effects  and   Higgs-boson
pole-mass shifts.   The code  {\tt CPsuperH} is  self--contained (with
all subroutines included),  is easy and fast to  run, and is organized
to   allow    further   theoretical   developments    to   be   easily
implemented\footnote{The     program    may    be     obtained    from
http://theory.ph.man.ac.uk/$^{{}_\sim}$jslee/CPsuperH.html}.  The fact
that the masses and couplings  of the charged and neutral Higgs bosons
are computed at a similar  high-precision level makes it an attractive
tool for Tevatron, LHC and LC studies, also in the CP-conserving case.

\newpage

\section{Introduction}

The quest for the still-elusive Higgs boson~\cite{Higgs}, the missing
cornerstone of the renormalizable Standard Model (SM), has become not
only more pressing, after the completion of the LEP experimental
programme, but also more exciting in light of the upcoming
experiments at the upgraded Tevatron collider and the Large Hadron
Collider (LHC). Indeed, direct searches for possible
realizations of the mechanism of spontaneous electroweak symmetry
breaking within and beyond the SM
are expected to dominate the scene of particle-physics phenomenology
in the present and next decades.

One of the most theoretically appealing realizations of the Higgs
mechanism for mass generation is provided by Supersymmetry~(SUSY).
The minimal supersymmetric extension of the SM (MSSM) has a number of
interesting field--theoretic and phenomenological properties, if SUSY
is softly broken such that superparticles acquire masses not greatly
exceeding 1 TeV. Specifically, within the MSSM, the gauge hierarchy
can be made technically natural~\cite{hierarchy,HPN}. Unlike the SM,
the MSSM exhibits quantitatively reliable gauge-coupling unification
at the energy scale of the order of $10^{16}$~GeV~\cite{DG}.
Furthermore, the MSSM provides a successful mechanism for cosmological
baryogenesis via a strongly first-order electroweak phase
transition~\cite{KRS,EWBAU}, and provides viable candidates for cold
dark matter~\cite{DM,DM2}.

The MSSM makes a crucial and definite prediction for future
high-energy experiments, that can be directly tested at the Tevatron
and/or the LHC. It guarantees the existence of (at least) one light
neutral Higgs boson with mass less than about 135~GeV~\cite{Mh}. This
rather strict upper bound on the lightest Higgs boson mass is in
accord with global analyses of the electroweak precision data, which
point towards a relatively light SM Higgs boson, with $M_{H_{\rm SM}}
\stackrel{<}{{}_\sim} 211$~GeV at the 95 \% confidence
level~\cite{LEPEWWG}. Furthermore, because of the decoupling
properties of heavy superpartners, the MSSM predictions for the
electroweak precision observables can easily be made consistent with
all the experimental data~\cite{ChoHagi}.

Recently, a new important phenomenological feature of the MSSM Higgs
sector has been observed. It has been realized that loop effects
mediated dominantly by third-generation squarks may lead to sizeable
violations of the tree-level CP invariance of the MSSM Higgs
potential, giving rise to significant Higgs scalar-pseudoscalar
transitions~\cite{APLB}, in particular. As a consequence, the three
neutral Higgs mass eigenstates $H_{1,2,3}$, labelled in order of
increasing mass such that $M_{H_1}\le M_{H_2} \le M_{H_3}$, have no
definite CP parities, but become mixtures of CP-even and CP-odd
states. Much work has been devoted to studying in greater detail this
radiative Higgs--sector CP violation in the framework of the
MSSM~\cite{PW,Demir,CDL,CEPW,KW,CHL,INhiggs,CPX,HeinCP,CEPW2,CEMPW}.
In the MSSM with explicit CP violation, the couplings of the Higgs
bosons to the SM gauge bosons and fermions, to their supersymmetric
partners and to the Higgs bosons themselves may be considerably
modified from those predicted in the CP-conserving case. Consequently,
radiative CP violation in the MSSM Higgs sector can significantly
affect the production rates and decay branching fractions of the Higgs
bosons. In particular, the drastic modification of the couplings of
the $Z$ boson to the two lighter Higgs bosons $H_1$ and $H_2$ might
enable a relatively light Higgs boson with a mass $M_{H_1}$ even less
than about 70~GeV to have escaped detection at
LEP~2~\cite{CEPW,CEMPW}. The upgraded Tevatron collider and the LHC
will be able to cover a large fraction of the MSSM parameter space,
including the challenging regions with a light Higgs boson without
definite CP parity~\cite{CPpp,CEMPW,CFLMP}. Furthermore, complementary
and more accurate explorations of the CP-noninvariant MSSM Higgs
sector can be carried out using high-luminosity $e^+e^-$~\cite{CPee}
and/or $\gamma\gamma$~\cite{CPphoton} colliders. In addition, a
complete determination of the CP properties of the neutral Higgs
bosons is possible at muon colliders by exploiting
polarized muon beams~\cite{CPmumu}.

It is obvious that a systematic study of Higgs phenomenology in
the MSSM with explicit CP violation would be greatly facilitated by
an appropriate computational tool. For this purpose, we have developed
the Fortran program {\tt CPsuperH}, a new self--contained computational
package, which calculates the mass spectrum and the decay widths of
the neutral and charged Higgs bosons in the MSSM with explicit CP
violation\footnote{We note in passing that a Fortran code called
{\tt HDECAY} has already been developed for the calculation of Higgs boson
decays in the CP--invariant version of the MSSM~\cite{HDECAY}.}. It
calculates the neutral Higgs-boson masses
$M_{H_{1,2,3}}$ and the corresponding $3\times 3$ Higgs-boson mixing
matrix $O$, employing the renormalization-group- (RG-)improved 
diagrammatic
approach of~\cite{CEPW2}.  We include the leading two-loop QCD
logarithmic corrections as well as the leading two-loop logarithmic
corrections induced by the top- and bottom-quark Yukawa
couplings~\cite{CQW}.  We also include the leading one-loop
logarithmic corrections due to gaugino and higgsino quantum
effects~\cite{HH}.  Moreover, we implement the potentially large
two--loop non--logarithmic corrections originating from one--loop
threshold effects on the top- and bottom-quark Yukawa couplings,
associated with the decoupling of the third-generation
squarks~\cite{CHHHWW,CEPW,CDL}.  Finally, the RG-improved diagrammatic
calculation takes account of mass shifts determined by the
positions of the poles in the corresponding Higgs-boson propagators.
These Higgs--boson pole-mass shifts for the lightest Higgs boson are
small, of the order of a few GeV. However, the mass shifts for the
heavier Higgs bosons, $H_2$ and $H_3$, can be much larger and of the
order of several tens of GeV~\cite{CEPW2}, especially if their masses
$M_{H_2}$ and $M_{H_3}$ happen to be close to thresholds for the
on-shell production of squark pairs. Finally, we note that the
computation of all the
Higgs-boson decay widths by the code {\tt CPsuperH} relies on the
extensive analytic results for the decay widths presented in~\cite{CHL}.

After this introductory discussion, the next section summarizes all
the topics of Higgs phenomenology that can be studied with the code
{\tt CPsuperH}.  In Section~\ref{sec:code}, we describe the execution
procedure of the Fortran code and present examples of input and output
files from a test run.  Finally, we summarize the essential
features of the code and provide an outlook for further developments
of {\tt CPsuperH} in Section~\ref{sec:summary}.

\section{Higgs Phenomenology in the MSSM with Explicit\\
  CP Violation}\label{sec:pheno}

In the presence of nontrivial CP--violating phases for the higgsino
mass parameter $\mu$ and the soft SUSY--breaking parameters in the
MSSM, the couplings of the Higgs bosons to the gauge bosons, fermions
and sfermions, as well as those to the Higgs bosons themselves, are
strongly modified. In order to investigate these modifications and
their phenomenological implications, we begin by stating our conventions
for the mixing matrices of the neutral Higgs bosons and SUSY particles,
and
then we present all the relevant Higgs interactions with the MSSM
particles to be used subsequently for calculating the masses, total
decay widths and decay branching fractions of the neutral and charged
Higgs bosons.

\subsection{Conventions}

In this subsection, we give our conventions for the mixing matrices of
neutral Higgs bosons, charginos, neutralinos and third--generation sfermions.
\begin{itemize}
\item \underline{Neutral Higgs bosons}: In the presence of
nontrivial CP-violating
phases of the soft supersymmetry-breaking parameters,
most relevantly in the
third-generation sfermion sector,
the three
neutral Higgs bosons all mix together via radiative corrections:
\begin{equation}
(\phi_1,\phi_2,a)^T_\alpha=O_{\alpha i}(H_1,H_2,H_3)^T_i \,,
\end{equation}
where $O^T{\cal M}^2_H\, O={\sf diag} \, (M^2_{H_1}, M^2_{H_2},
M^2_{H_3})$ with $M_{H_1} \leq M_{H_2}\leq M_{H_3}$.
We refer to~\cite{CEPW,CEPW2} for the details of the calculations of the
mass-squared matrix ${\cal M}^2_H$, the diagonalization matrix $O$ and
the pole masses of the Higgs bosons.
\item \underline{Charginos}:
In SUSY theories, the spin--1/2 partners of the $W^\pm$
gauge bosons and the charged Higgs bosons, $\tilde{W}^\pm$ and
$\tilde{H}^\pm$,
mix to form chargino mass eigenstates. We adopt the convention
$\tilde{H}^{-}_{L(R)} = \tilde{H}^{-}_{1(2)}$, where the subscripts
1 and 2 are associated with the Higgs supermultiplets leading to the tree-level
mass generation of the down- and up-type quarks, respectively.
The chargino mass matrix in the $(\tilde{W}^-,\tilde{H}^-)$ basis
\begin{eqnarray}
{\cal M}_C = \left(\begin{array}{cc}
     M_2              & \sqrt{2} M_W\, c_{\beta} \\[2mm]
\sqrt{2} M_W\, s_{\beta} & \mu
             \end{array}\right)\, ,
\end{eqnarray}
is diagonalized by two different unitary matrices
$ C_R{\cal M}_C C_L^\dagger ={\sf diag}\{m_{\tilde{\chi}^\pm_1},\,
m_{\tilde{\chi}^\pm_2}\}$, where
$m_{\tilde{\chi}^\pm_1} \leq m_{\tilde{\chi}^\pm_2}$.
The chargino mixing matrices $(C_L)_{i\alpha}$ and $(C_R)_{i\alpha}$
relate the electroweak eigenstates to the mass eigenstates, via
\begin{eqnarray}
\tilde{\chi}^-_{\alpha L} &=&
(C_L)^*_{i \alpha } \tilde{\chi}_{iL}^-\,,\qquad
\tilde{\chi}^-_{\alpha L}\ =\ (\tilde{W}^-, \tilde{H}^-)_L^T\,,\nonumber\\
\tilde{\chi}^-_{\alpha R} &=& (C_R)^*_{i \alpha} \tilde{\chi}_{iR}^-\,,\qquad
\tilde{\chi}^-_{\alpha R}\ =\ (\tilde{W}^-, \tilde{H}^-)_R^T\,.
\end{eqnarray}
We use the following abbreviations throughout this paper:
$s_\beta\equiv\sin\beta$, $c_\beta\equiv\cos\beta$,
$t_\beta=\tan\beta$, $s_{2\beta}\equiv\sin\,2\beta$,
$c_{2\beta}\equiv\cos\,2\beta$, $s_W\equiv\sin\theta_W$,
$c_W\equiv\cos\theta_W$, etc.
\item \underline{Neutralinos}:
The neutralino mass matrix in the
$(\tilde{B},\tilde{W}^3,\tilde{H}^0_1,\tilde{H}^0_2)$ basis is given by
\begin{eqnarray}
{\cal M}_N=\left(\begin{array}{cccc}
  M_1       &      0          &  -M_Z c_\beta s_W  & M_Z s_\beta s_W \\[2mm]
   0        &     M_2         &   M_Z c_\beta c_W  & -M_Z s_\beta c_W\\[2mm]
-M_Z c_\beta s_W & M_Z c_\beta c_W &       0       &     -\mu        \\[2mm]
 M_Z s_\beta s_W &-M_Z s_\beta c_W &     -\mu      &       0
                  \end{array}\right)\,.
\end{eqnarray}
This neutralino mass matrix is diagonalized by a unitary matrix $N$:
$N^* {\cal M}_N N^\dagger = {\sf diag}\,
(m_{\tilde{\chi}_1^0},m_{\tilde{\chi}_2^0},m_{\tilde{\chi}_3^0},m_{\tilde{\chi}_4^0})$
with
$m_{\tilde{\chi}_1^0} \leq m_{\tilde{\chi}_2^0} \leq m_{\tilde{\chi}_3^0}
\leq m_{\tilde{\chi}_4^0}$.
The neutralino mixing matrix $N_{i\alpha}$
relates the electroweak eigenstates to the mass eigenstates via
\begin{eqnarray}
(\tilde{B},\tilde{W}^3,\tilde{H}^0_1,\tilde{H}^0_2)^T_{\alpha}
=N_{i\alpha}^*
({\tilde{\chi}_1^0},{\tilde{\chi}_2^0},{\tilde{\chi}_3^0},
{\tilde{\chi}_4^0})^T_{i}\,.
\end{eqnarray}
%
\item \underline{Stops, sbottoms, staus and tau sneutrino}: The stop and sbottom mass
matrices may conveniently be written in the $\left(\tilde{q}_L,
\tilde{q}_R\right)$ basis as
\begin{equation}
  \label{Mscalar}
\hspace{-0.5 cm}
\widetilde{\cal M}^2_q  = \left( \begin{array}{cc}
M^2_{\tilde{Q}_3}\, +\, m^2_q\, +\, c_{2\beta} M^2_Z\, ( T^q_z\, -\,
Q_q s_W^2 ) & h_q^* v_q (A^*_q - \mu R_q )/\sqrt{2}\\
h_q v_q (A_q - \mu^* R_q)/\sqrt{2} & \hspace{-0.2cm}
M^2_{\tilde{R}_3}\, +\, m^2_q\, +\, c_{2\beta} M^2_Z\, Q_q s^2_W
\end{array}\right)\, ,
\end{equation}
with
$q=t,b$, $R=U,D$,
$T^t_z = - T^b_z = 1/2$,
$Q_t = 2/3$, $Q_b = -1/3$,
$v_b=v_1$, $v_t=v_2$,
$R_b = \tan\beta = v_2/v_1$, $R_t = \cot\beta$, and
$h_q$ is the Yukawa coupling of the quark $q$.
On the other hand, the stau mass matrix is written in the $\left(\tilde{\tau}_L,
\tilde{\tau}_R\right)$ basis as
\begin{eqnarray}
  \label{Mstau}
\hspace{-0.5 cm}
\widetilde{\cal M}^2_\tau  = \left( \begin{array}{cc}
M^2_{\tilde{L}_3}\, +\, m^2_\tau\, +\, c_{2\beta} M^2_Z\, (s_W^2-1/2 )
   & h_\tau^* v_1 (A^*_\tau - \mu \tan\beta )/\sqrt{2}\\
h_\tau v_1 (A_\tau - \mu^* \tan\beta)/\sqrt{2} & \hspace{-0.2cm}
M^2_{\tilde{E}_3}\, +\, m^2_\tau\, -\, c_{2\beta} M^2_Z\, s^2_W
\end{array}\right)\, ,
\end{eqnarray}
and the mass of the tau sneutrino $\tilde{\nu_\tau}$ is simply
$m_{\tilde\nu_\tau} = \sqrt{ M^2_{\tilde{L}_3} + \frac{1}{2}
c_{2\beta} M_Z^2 }$, as it has no right--handed counterpart in the
MSSM.  The $2\times 2$ sfermion mass matrix $\widetilde{\cal M}^2_f$
for $f=t, b$ and $\tau$ is diagonalized by a unitary matrix
$U^{\tilde{f}}$: $U^{\tilde{f}\dagger} \, \widetilde{\cal M}^2_f \,
U^{\tilde{f}} ={\sf diag}(m_{\tilde{f}_1}^2,m_{\tilde{f}_2}^2)\,$ with
$m_{\tilde{f}_1}^2 \leq m_{\tilde{f}_2}^2$.  The mixing matrix
$U^{\tilde{f}}$ relates the electroweak eigenstates $\tilde{f}_{L,R}$
to the mass eigenstates $\tilde{f}_{1,2}$, via
\begin{equation}
(\tilde{f}_L,\tilde{f}_R)^T_\alpha\,=\,
U^{\tilde{f}}_{\alpha i} \,
(\tilde{f}_1,\tilde{f}_2)^T_i\,.
\end{equation}
We parameterize the mixing matrices as follows:
\begin{eqnarray}
U^{\tilde{f}}=
\left( \begin{array}{cc}
\cos\theta_{\tilde{f}} & -\sin\theta_{\tilde{f}}\, {\rm e}^{-i\phi_{\tilde{f}}}\\
\sin\theta_{\tilde{f}}\, {\rm e}^{+i\phi_{\tilde{f}}}& \cos\theta_{\tilde{f}}
       \end{array}
\right)\,,
\end{eqnarray}
and we calculate numerically the mixing angle $\theta_{\tilde{f}}$ and 
phase
$\phi_{\tilde{f}}$ in the ranges between $-\pi/2$ and $\pi/2$, so that
$\cos\theta_{\tilde{f}}\geq 0$ and
$\cos\phi_{\tilde{f}}\geq 0$.
\end{itemize}

\subsection{Higgs--Boson Interactions}

In this subsection, we list all the Higgs interactions with gauge
bosons,
SM fermions, squarks, sleptons, charginos, and neutralinos. We also
present all the trilinear and quartic Higgs--boson self--couplings.
\begin{itemize}
\item \underline{Interactions of Higgs bosons with gauge bosons}: The
interactions of the Higgs bosons with the gauge bosons $Z$ and $W^\pm$
are described by the three interaction Lagrangians:
\begin{eqnarray}
{\cal L}_{HVV} & = & g\,M_W \, \left(W^+_\mu W^{- \mu}\ + \
\frac{1}{2c_W^2}\,Z_\mu Z^\mu\right) \, \sum_i \,g_{_{H_iVV}}\, H_i
\,,\\[3mm]
{\cal L}_{HHZ} &=& \frac{g}{4c_W} \sum_{i,j} g_{_{H_iH_jZ}}\, Z^{\mu}
(H_i\, i\!\stackrel {\leftrightarrow} {\partial}_\mu H_j) \,, \\ [3mm]
{\cal L}_{HH^\pm W^\mp} &=& -\frac{g}{2} \, \sum_i \, g_{_{H_iH^+
W^-}}\, W^{-\mu} (H_i\, i\!\stackrel{\leftrightarrow}{\partial}_\mu
H^+)\, +\, {\rm h.c.}\,,
\end{eqnarray}
where $g = e/\sin\theta_W$ is the SU(2)$_L$ gauge-coupling constant,
and  the couplings $g_{_{H_iVV}}$, $g_{_{H_iH_jZ}}$ and $g_{_{H_iH^+
W^-}}$ are given in terms of the neutral Higgs-boson mixing matrix $O$
by (note that det$(O)=\pm1$ for any orthogonal matrix $O$):
\begin{eqnarray}
g_{_{H_iVV}} &=& c_\beta\, O_{\phi_1 i}\: +\: s_\beta\, O_{\phi_2 i}
\, ,\nonumber \\
g_{_{H_iH_jZ}} &=& {\rm sign} [{\rm det}(O)] \, \, \varepsilon_{ijk}\,
g_{_{H_kVV}}\,,  \nonumber \\
g_{_{H_iH^+ W^-}} &=& c_\beta\, O_{\phi_2 i} - s_\beta\, O_{\phi_1 i}
- i O_{ai} \, ,
\end{eqnarray}
leading to the following sum rules:
\begin{equation}
\sum_{i=1}^3\, g_{_{H_iVV}}^2\ =\ 1\,\quad{\rm and}\quad
g_{_{H_iVV}}^2+|g_{_{H_iH^+ W^-}}|^2\ =\ 1\,\quad {\rm for~ each}~ i\,.
\end{equation}
%

\item \underline{Higgs--quark--antiquark and Higgs--lepton-antilepton
interactions}:
The effective Lagrangian governing the interactions of the neutral
Higgs bosons with quarks and charged leptons is
\begin{equation}
{\cal L}_{H_i\bar{f}f}\ =\ - \sum_{f=u,d,l}\,\frac{g m_f}{2 M_W}\,
\sum_{i=1}^3\, H_i\, \bar{f}\,\Big( g^S_{H_i\bar{f}f}\, +\,
ig^P_{H_i\bar{f}f}\gamma_5 \Big)\, f\ .
\end{equation}
At the tree level, $(g^S,g^P) = (O_{\phi_1 i}/c_\beta, -O_{ai}
\tan\beta)$ and $(g^S, g^P) = (O_{\phi_2 i}/s_\beta, -O_{ai}
\cot\beta)$ for $f=(l,d)$ and $f=u$, respectively. In the case of
third-generation quarks, the program {\tt CPsuperH} computes the finite
threshold
corrections induced by the exchanges of gluinos and charginos. As
described in Appendix A, we include the all--orders
resummation~\cite{resum1,EHKMY,CEMPW} of the leading powers of $\tan\beta$, as
required for a meaningful perturbative expansion. Correspondingly, in
the presence of CP violation, the effective couplings of the charged
Higgs boson to quarks and leptons in the weak-interaction basis are
described by the interaction Lagrangian:
\begin{equation} \label{Hud}
{\cal L}_{H^\pm f_{\uparrow}f_{\downarrow}}\ =\ \frac{g}{\sqrt{2}
M_W}\, \hskip -0.2cm
\sum_{(f_{\uparrow},f_{\downarrow})=(u,d),(\nu,l)} \hskip -0.3cm
H^+\, \bar{f}_{\uparrow}\,\Big(\, m_{f_{\uparrow}}\,
g^L_{H^+\bar{f}_{\uparrow}f_{\downarrow}}\,
P_L\ +\ m_{f_{\downarrow}}\,
g^R_{H^+\bar{f}_{\uparrow}f_{\downarrow}}\, P_R\, \Big)\, f_{\downarrow}\ +\ {\rm h.c.} ,
\end{equation}
where $P_{L/R} \equiv (1\, \mp\, \gamma_5)/2$. At the tree level,
$g^L=\cot\beta$ and $g^R=\tan\beta$ with $m_\nu=0$. The loop--induced
threshold corrections to the $g^{L,R}_{H^+\bar{u}d }$ couplings are
presented in Appendix A.

\item \underline{Higgs--sfermion--sfermion interactions}:
The Higgs--sfermion--sfermion interactions can be written in terms of the
sfermion mass eigenstates as
\begin{equation}
{\cal L}_{H\tilde{f}\tilde{f}}=v\sum_{f=u,d}\,g_{H_i\tilde{f}^*_j\tilde{f}_k}
(H_i\,\tilde{f}^*_j\,\tilde{f}_k)\,,
\end{equation}
where
$$ v\,g_{H_i\tilde{f}^*_j\tilde{f}_k}
=\left(\Gamma^{\alpha\tilde{f}^*\tilde{f}}\right)_{\beta\gamma}
O_{\alpha i}U^{\tilde{f}*}_{\beta j} U^{\tilde{f}}_{\gamma k}\,,
$$
with $\alpha=(\phi_1,\phi_2,a)=(1,2,3)$, $\beta,\gamma = L, R$,
$i=(H_1,H_2,H_3)=(1,2,3)$ and $j,k=1,2$. Likewise, the charged Higgs-boson
interactions with up-- and down--type sfermions are given by
\begin{equation}
{\cal L}_{H^\pm\tilde{f}\tilde{f'}}=v\,g_{H^+\tilde{f}^*_j\tilde{f'}_k}
(H^+\,\tilde{f}^*_j\,\tilde{f'}_k) + {\rm h.c.},
\end{equation}
where
$$
v\, g_{H^+\tilde{f}^*_j\tilde{f'}_k}\ =\
\left(\Gamma^{H^+\tilde{f}^*\tilde{f'}}\right)_{\beta\gamma}\,
U^{\tilde{f}*}_{\beta j}\, U^{\tilde{f'}}_{\gamma k}\,.
$$
The expressions for the couplings
$\Gamma^{\alpha\tilde{f}^*\tilde{f}}$ and
$\Gamma^{H^+\tilde{f}^*\tilde{f'}}$ to third--generation sfermions are
presented in Appendix B. As shown in~\cite{EHKMY}, our
iterative treatment of the threshold corrections that are enhanced at large
$\tan\beta$ ensures that the corresponding corrections to the Higgs-boson 
couplings to squarks are also resummed correspondingly.

\item \underline{Interactions of  neutral Higgs bosons and charginos}:
These are described by the following Lagrangian:
\begin{eqnarray}
{\cal L}_{H^0\wt{\chi}^+\wt{\chi}^-}
&=&-\frac{g}{\sqrt{2}}\sum_{i,j,k} H_k
\overline{\wt{\chi}_i^-}
\left(g_{H_k\tilde{\chi}^+_i\tilde{\chi}^-_j}^{S}+i\gamma_5
g_{H_k\tilde{\chi}^+_i\tilde{\chi}^-_j}^{P}\right)
\wt{\chi}_j^-\,,
\nonumber \\
g_{H_k\tilde{\chi}^+_i\tilde{\chi}^-_j}^{S}&=&\frac{1}{2}\left\{
[(C_R)_{i1}(C_L)^*_{j2}G^{\phi_1}_k+(C_R)_{i2}(C_L)^*_{j1}G^{\phi_2}_k]
+[i\leftrightarrow j]^* \right\}\,,
\nonumber \\
g_{H_k\tilde{\chi}^+_i\tilde{\chi}^-_j}^{P} &=&\frac{i}{2}\left\{
[(C_R)_{i1}(C_L)^*_{j2}G^{\phi_1}_k+(C_R)_{i2}(C_L)^*_{j1}G^{\phi_2}_k]
-[i\leftrightarrow j]^* \right\}\,,
\end{eqnarray}
where $G^{\phi_1}_k=(O_{\phi_1 k}-is_\beta O_{ak})$,
$G^{\phi_2}_k=(O_{\phi_2 k}-ic_\beta O_{ak})$,
$i,j=1,2$, and $k=1-3$.
\item \underline{Interactions of neutral Higgs bosons and neutralinos}:
These are described by the following Lagrangian:
\begin{eqnarray}
{\cal L}_{H^0\wt{\chi}^0\wt{\chi}^0}
&=&-\frac{g}{2}\sum_{i,j,k} H_k
\overline{\wt{\chi}_i^0}
\left(g_{H_k\tilde{\chi}^0_i\tilde{\chi}^0_j}^{S}
+i\gamma_5
g_{H_k\tilde{\chi}^0_i\tilde{\chi}^0_j}^{P}\right)
\wt{\chi}_j^0 \,:
\nonumber \\
g_{H_k\tilde{\chi}^0_i\tilde{\chi}^0_j}^{S}
&=&\frac{1}{2}\real{[(N_{j2}^*-t_W N_{j1}^*)
(N_{i3}^*G^{\phi_1}_k-N_{i4}^*G^{\phi_2}_k)+(i\leftrightarrow j)]}
\nonumber \\
g_{H_k\tilde{\chi}^0_i\tilde{\chi}^0_j}^{P}
&=&-\frac{1}{2}\imag{[(N_{j2}^*-t_W N_{j1}^*)
(N_{i3}^*G^{\phi_1}_k-N_{i4}^*G^{\phi_2}_k)+(i\leftrightarrow j)]}\,,
\end{eqnarray}
where $i,j=1$-$4$ for the four neutralino states and $k=1$-$3$ for
the three neutral Higgs bosons.
\item \underline{Interactions of charged Higgs bosons, charginos and neutralinos}:
These are described by the following Lagrangian:
\begin{eqnarray}
{\cal L}_{H^\pm\wt{\chi}_i^0\wt{\chi}_j^\mp}
&=&-\frac{g}{\sqrt{2}}\sum_{i,j} H^+
\,\overline{\wt{\chi}_i^0}
\left(g_{H^+\tilde{\chi}^0_i\tilde{\chi}^-_j}^{S}
+i\gamma_5
g_{H^+\tilde{\chi}^0_i\tilde{\chi}^-_j}^{P}\right)
\wt{\chi}_j^-
+{\rm h.c.}\,: \nonumber \\
g_{H^+\tilde{\chi}^0_i\tilde{\chi}^-_j}^{S}
&=&\frac{1}{2}\left\{s_\beta\left[\sqrt{2}N_{i3}^*(C_L)_{j1}^*-(N_{i2}^*
+t_W N_{i1}^*)(C_L)_{j2}^*\right] \right. \nonumber \\
&&\left. ~  +c_\beta\left[\sqrt{2}N_{i4}(C_R)_{j1}^*+(N_{i2}
+t_W N_{i1})(C_R)_{j2}^*\right]\right\} \,, \nonumber \\
g_{H^+\tilde{\chi}^0_i\tilde{\chi}^-_j}^{P}
&=&\frac{i}{2}\left\{s_\beta\left[\sqrt{2}N_{i3}^*(C_L)_{j1}^*-(N_{i2}^*
+t_W N_{i1}^*)(C_L)_{j2}^*\right] \right. \nonumber \\
&&\left. ~  -c_\beta\left[\sqrt{2}N_{i4}(C_R)_{j1}^*+(N_{i2}
+t_W N_{i1})(C_R)_{j2}^*\right]\right\} \,.
\end{eqnarray}

\item \underline{Trilinear and quartic Higgs-boson self-couplings
~\cite{CHL,CEMPW}}: The effective trilinear and quartic Higgs self--couplings
can be cast into the form
\begin{eqnarray} \label{hselfcoup}
{\cal L}_{3H} & = &v\,\sum_{i\geq j\geq k=1}^3 g_{_{H_iH_jH_k}}\, H_i
  H_j H_k\: +\: v\,\sum_{i=1}^3 g_{_{H_iH^+H^-}}\,H_iH^+H^-\,,\\[3mm]
{\cal L}_{4H} & =& \sum_{i\geq j\geq k\geq l =1}^3
  g_{_{H_iH_jH_kH_l}}\, H_i H_j H_k H_l\
+\ \sum_{i\geq j =1}^3 g_{_{H_iH_j H^+H^-}}\, H_i H_j H^+
  H^-\nonumber\\
&&+\  g_{_{H^+H^-H^+H^-}}\, (H^+ H^-)^2\,,\
\end{eqnarray}
where
\begin{eqnarray}
g_{_{H_iH_jH_k}} &=& \sum_{\alpha\leq \beta\leq \gamma=1}^3
\left\{O_{\alpha i} O_{\beta j} O_{\gamma  k} \right\}\,
g_{\alpha\beta\gamma} \,, \quad
g_{_{H_iH^+H^-}}\ =\ \sum_{\alpha =1}^3 O_{\alpha i}\,
g_{_{\alpha H^+H^-}}\,,\quad
\label{eq:22}\\[3mm]
g_{_{H_iH_jH_kH_l}} &=&  \sum_{\alpha\leq \beta\leq \gamma \leq \delta=1}^3
\left\{O_{\alpha i} O_{\beta j} O_{\gamma  k} O_{\delta  l} \right\}\,
g_{\alpha\beta\gamma\delta} \,, \nonumber\\[3mm]
g_{_{H_iH_j H^+H^-}} & =& \sum_{\alpha\leq \beta =1}^3
\left\{O_{\alpha i} O_{\beta j} \right\}\,
g_{\alpha\beta {\scriptscriptstyle H^+H^-}}\,.
\label{eq:23}
\end{eqnarray}
In the above equations (\ref{eq:22}) and (\ref{eq:23}), the expressions within
the curly brackets $\{ \cdots \}$ need to be symmetrized with respect to
the indices $i,j,k,l$ and divided by the corresponding symmetry factors in
cases where
two or more indices are the same. For example, $\left\{ O_{\alpha i} O_{\beta j} O_{\gamma
k} \right\}$ can explicitly be evaluated as follows:
\begin{eqnarray}
\left\{   O_{\alpha i} O_{\beta j} O_{\gamma  k} \right\} &\equiv&
      \frac{1}{N_S}\, \Big(
          O_{\alpha i} O_{\beta j} O_{\gamma  k}
         +O_{\alpha i} O_{\beta k} O_{\gamma  j}
         +O_{\alpha j} O_{\beta i} O_{\gamma  k}
         +O_{\alpha j} O_{\beta k} O_{\gamma  i}  \nonumber \\
&&       +O_{\alpha k} O_{\beta i} O_{\gamma  j}
         +O_{\alpha k} O_{\beta j} O_{\gamma  i}\Big)\,,
\end{eqnarray}
with $N_S=6$ when $i=j=k$, $N_S=1$ when $(i,j,k)=(3,2,1)$, and $N_S=2$ in all
the other cases. We present the couplings
$g_{\alpha\beta\gamma}$, $g_{\alpha H^+H^-}$,
$g_{\alpha\beta\gamma\delta}$, and $g_{\alpha\beta H^+H^-}$
of the Higgs    weak eigenstates in Appendix C.
\end{itemize}
%

\subsection{Neutral and Charged Higgs Boson Decays}

In this subsection, we calculate all the two--body decay widths of the
Higgs bosons.  We consider the decays of the Higgs bosons into pairs
of leptons, quarks, charginos, neutralinos, massive gauge bosons,
Higgs bosons, squarks, sleptons, photons and gluons as well as into a massive
gauge boson and a Higgs boson. For the decay modes involving more than
one massive gauge boson, three-body decays are also considered~\cite{CHL}.

\begin{itemize}
\item \underline{$H_i,H^+\rightarrow f f^\prime$}:
First, let us consider the decays
into a pair of fermions. Without loss of generality, the Lagrangian describing
the interactions of the Higgs bosons with two fermions can be written as
\begin{eqnarray}
\label{eq:hff}
{\cal L}_{Hff^\prime}&=&-\sum_{i,j,k} g_f H_i \bar{f}_k
(g^S_{_{H_i\bar{f}_kf_j}}+ig^P_{_{H_i\bar{f}_kf_j}}\gamma_5)f_j \nonumber \\
&&-\left[\sum_{j,k} g_{ff^\prime} H^{+} \bar{f}_k
(g^S_{_{H^+\bar{f}_kf_j^\prime}}+ig^P_{_{H^+\bar{f}_kf_j^\prime}}\gamma_5)
f_j ^\prime
\ \ +{\rm h.c.}\right]\,,
\end{eqnarray}
where $f^{(\prime)}$ stands for a lepton, a quark, a chargino, or a
neutralino, and
the tree--level couplings $g_{f}$, $g_{ff^\prime}$ and $g^{S,P}$ are given
in Table \ref{tab:hff}.
In terms of these generic couplings,
the width for a decay into two Dirac fermions is given by
\begin{equation}
\hspace{-0.5 cm}
\Gamma_D = N_C
\frac{g_{f(f^\prime)}^2 M_H \lambda^{1/2}(1,\kappa_j,\kappa_k)}{8\pi}
\left[(1\!-\!\kappa_j\!-\!\kappa_k)(|g^S|^2+|g^P|^2)
-2\sqrt{\kappa_j\kappa_k} (|g^S|^2-|g^P|^2)\right]\,,
\end{equation}
where $\kappa_j \equiv m_{f_j}^2/M_H^2$ and
$\lambda(1,x,y) \equiv (1-x-y)^2-4xy$. We note that
$\Gamma_D$ becomes
$N_C\frac{g_{f(f^\prime)}^2 M_H \beta_\kappa}{8\pi}
(\beta_k^2|g_S|^2+|g_P|^2)$  when $\kappa_j=\kappa_k=\kappa$, where
$\beta_\kappa \equiv \sqrt{1-4\kappa}$.
The colour factor
$N_C=3$ for quarks and 1 for leptons, charginos, and neutralinos.
The decay widths into two Majorana fermions are given by
\begin{equation}
\Gamma_M= \left(\frac{4}{1+\delta_{jk}}\right)\,\Gamma_D\,,
\end{equation}
where $\delta_{jk}=1$ for identical Majorana fermions.

\begin{table}[\hbt]
\caption{\label{tab:hff}
{\it
The couplings $g_f$, $g_{ff^\prime}$ and $g^{S,P}$ in Eq.~(\ref{eq:hff})
at the tree level.
}}
\begin{center}
\begin{tabular}{|l|ccc|}
\hline
Decay Mode  &  $g_f$  & $g^S$  & $g^P$    \\
\hline
$H_i\rightarrow l^+ l^-\,$  & $\frac{g m_l}{2 M_W}$
 & $O_{\phi_1 i}/c_\beta$ & $-(s_\beta/c_\beta)O_{a i}$ \\
$H_i\rightarrow d \bar{d}\,$  & $\frac{g m_d}{2 M_W}$
 & $O_{\phi_1 i}/c_\beta$  & $-(s_\beta/c_\beta)O_{a i}$ \\
$H_i\rightarrow u \bar{u}\,$   & $\frac{g m_u}{2 M_W}$
 & $O_{\phi_2 i}/s_\beta$  & $-(c_\beta/s_\beta)O_{a i}$ \\
$H_i\rightarrow \tilde{\chi}_j^0 \tilde{\chi}_k^0\,$
 & $g/2$
 & $g_{H_i\tilde{\chi}^0_j\tilde{\chi}^0_k}^{S}$
 & $g_{H_i\tilde{\chi}^0_j\tilde{\chi}^0_k}^{P}$ \\
$H_i\rightarrow \tilde{\chi}_j^- \tilde{\chi}_k^+\,$
  & $g/\sqrt{2}$
 & $g_{H_i\tilde{\chi}^+_j\tilde{\chi}^-_k}^{S}$
 & $g_{H_i\tilde{\chi}^+_j\tilde{\chi}^-_k}^{P}$ \\
\hline
Decay Mode  &  $g_{ff^\prime}$  & $g^S$  & $g^P$  \\
\hline
$H^+\rightarrow l^+ \nu\,$
& $-\frac{gm_l}{\sqrt{2}M_W}$
 & $t_\beta/2$
 & $-it_\beta/2$ \\
$H^+\rightarrow u \bar{d}\,$
& $-\frac{gm_u}{\sqrt{2}M_W}$
 & $[1/t_\beta+(m_d/m_u)\,t_\beta]/2$
 & $i[1/t_\beta-(m_d/m_u)\,t_\beta]/2$ \\
$H^+\rightarrow \tilde{\chi}_j^+\tilde{\chi}_i^0\,$
 & $g/\sqrt{2}$
 & $g_{H^+\tilde{\chi}^0_i\tilde{\chi}^-_j}^{S}$
 & $g_{H^+\tilde{\chi}^0_i\tilde{\chi}^-_j}^{P}$  \\
\hline
\end{tabular}
\end{center}
\end{table}
For the couplings $g^{S,P}_{H_i\bar{b}b}$, $g^{S,P}_{H_i\bar{t}t}$ and
$g^{S,P}_{H^+\bar{t}b}$, the finite loop--induced threshold
corrections due to the exchanges of gluinos and charginos can be
included by taking {\tt IFLAG\_H(10)}=0 (the default setting) in the
code {\tt CPsuperH}, as explained in Sec.~\ref{sec:code}.

\smallskip

For $H_i \rightarrow q\bar{q}$, 
the leading-order QCD correction is
taken into account by applying the enhancement factor
$K^q_i \equiv 1+5.67\,\frac{\alpha_s(M_{H_i}^2)}{\pi}$ to the decay width
given above. We take the running fermion masses
at the scale $m_t^{\rm pole}$ as reference values.
The effect on the couplings of the running of the quark masses from the
top--quark pole mass scale to the Higgs--boson mass scale is
also considered in calculating the corresponding decay widths, as
\begin{equation}
\Gamma(H_i\rightarrow q \bar{q}) = K^q_i
\left(\frac{m_q(M_{H_i})}{m_q(m_t^{\rm pole})}\right)^2
\Gamma_D(H_i\rightarrow q \bar{q}) \,,
\end{equation}
where $m_q(m_t^{\rm pole})$ is used in $g_f$, but $m_q(m_{H_i})$ is
used in $\kappa_f$, when calculating $\Gamma_D$. Likewise, running
$q$ and $q'$ quark masses are used when computing $\Gamma(H^+
\rightarrow q \bar{q'})$, while the dominant one-loop QCD corrections 
have been included by factors very similar to $K_i^q$.  
Finite quark-mass and higher-order QCD effects will depend, to some
extent, on the CP-violating parameters of the MSSM, and require an
independent study. In the present version of the code, we only include
the leading-order QCD effects which remain unaffected by CP violation.
Also, we do not include  flavour-violating decays of the neutral Higgs
bosons.  We  plan to  implement such refinements  in future  versions of
{\tt CPsuperH}.

\item \underline{$H_i\rightarrow VV$}: The width for decay into two
massive gauge bosons is given by
\begin{eqnarray}
\Gamma&=& \frac{G_F\,g_{_{H_iVV}}^2 M_{H_i}^3\,\delta_V}{16\sqrt{2}\pi}
\beta_{iV}(1-4\kappa_{iV}+12\kappa_{iV}^2)\,,
\end{eqnarray}
where $G_F/\sqrt{2}=g^2/8M_W^2$, $\kappa_{iV}=M_V^2/M_{H_i}^2$,
$\beta_{iV}=\sqrt{1-4\kappa_{iV}}$, and
$\delta_W=2$ and $\delta_Z=M_W^4/(c_W\,M_Z)^4=1$.
The three--body decay width
$\Gamma(H_i\rightarrow VV^*)$ is also calculated, using~\cite{CHL}:
\begin{equation}
\Gamma(H_i\rightarrow VV^*)=
\frac{G_Fg_{_{H_iVV}}^2 M_{H_i}^3\delta_V\delta_{VV^*}}{16\sqrt{2}\pi}
\int_{0}^{(\sqrt{\omega_i}-1)^2}{\rm d}x
\frac{\epsilon_V\lambda^{1/2}(\omega_i,x,1)[\lambda(\omega_i,x,1)+12x]}
{\omega_i^3\pi[(x-1)^2+\epsilon_V^2]}\,,
\end{equation}
where $\omega_i=M_{H_i}^2/M_V^2=1/\kappa_{iV}$, $\delta_{VV^*}=2$, and
$\epsilon_V=\Gamma_V/M_V$. We note that
$$
\lambda^{1/2}(\omega_i,1,1)[\lambda(\omega_i,1,1)+12]/\omega_i^3
=\beta_{iV}(1-4\kappa_{iV}+12\kappa_{iV}^2)\,.
$$
\item \underline{$H_i\rightarrow H_jZ$ and $H^{+}\rightarrow H_jW^+$}:
The decay width of a heavier Higgs boson into a lighter
Higgs boson and a massive gauge boson is given by
\begin{equation}
\Gamma \ = \
\frac{G_F M_H^3}{8\sqrt{2}\pi} |{\cal G}|^2
\lambda^{3/2}(1,\kappa_j,\kappa_V)\,,
\end{equation}
where $(M_H,{\cal G}) = (M_{H_i},g_{_{H_iH_jZ}})$ or
$(M_{H^+},g_{_{H_jH^+ W^-}})$, $\kappa_j=M_{H_j}^2/M_{H}^2$,
and $\kappa_V=M_V^2/M_{H}^2$.
The three--body decay widths
$\Gamma(H_i\rightarrow H_j Z^*)$  and
$\Gamma(H^+\rightarrow H_j W^{+*})$
are given~\cite{CHL} by:
\begin{equation}
\Gamma(H_i\rightarrow H_j Z^*) =
\frac{G_FM_{H_i}^3g_{_{H_iH_jZ}}^2}{8\sqrt{2}\pi}
\int_0^{(\sqrt{\omega_i}-\sqrt{\omega_j})^2}{\rm d}x
\frac{\epsilon_Z\lambda^{3/2}(\omega_i,\omega_j,x)}
{\omega_i^3\pi[(x-1)^2+\epsilon_Z^2]}\,,
\end{equation}
with $\omega_i=M_{H_i}^2/M_Z^2$, and similarly
\begin{equation}
\Gamma(H^+\rightarrow H_j W^{+*}) =
\frac{G_FM_{H^\pm}^3|g_{_{H_jH^+W^-}}|^2}{8\sqrt{2}\pi}
\int_0^{(\sqrt{\omega_\pm}-\sqrt{\omega_j})^2}{\rm d}x
\frac{\epsilon_W\lambda^{3/2}(\omega_\pm,\omega_j,x)}
{\omega_\pm^3\pi[(x-1)^2+\epsilon_W^2]}\,,
\end{equation}
with $\omega_j=M_{H_j}^2/M_W^2$ and
$\omega_\pm=M_{H^\pm}^2/M_W^2$.
\item \underline{$H_i\rightarrow H_jH_k$, $H_{i}\rightarrow
\tilde{f}_j\widetilde{f}_k^*$, and $H^+\rightarrow
\tilde{f}_j\widetilde{f'}^*_k$}: The decay widths into two scalar
particles can be written as
\begin{equation}
\Gamma\ =\ N_F\frac{v^2 |{\cal G}|^2}{16\pi
M_{H_i}}\lambda^{1/2}(1,\kappa_j,\kappa_k)\,,
\end{equation}
where\footnote{We note that the couplings $g_{_{H_iH_jH_k}}$ are defined
via the Lagrangian (\ref{hselfcoup}). The $H_i H_j H_j$ vertices for
$i > j$ therefore contain an extra factor of 2.} $(N_F,{\cal G}) =
(1+\delta_{jk},g_{_{H_iH_jH_k}})$, $(N_C,
g_{H_i\tilde{f}^*_j\tilde{f}_k})$ or $(N_C,
g_{H^+\tilde{f}^*_j\tilde{f'}_k})$, and
$\kappa_j=M_{H_j,\tilde{f_j}}^2/M_{H_i}^2$.
%
\item \underline{$H_i\rightarrow \gamma\gamma$}:
The amplitude for the decay process
$H_i \rightarrow \gamma\gamma$ can be written as
\begin{eqnarray} \label{hipp}
{\cal M}_{\gamma\gamma H_i}=-\frac{\alpha M_{H_i}^2}{4\pi\,v}
\bigg\{S^\gamma_i(M_{H_i})\, \left(\epsilon^*_{1\perp}\cdot\epsilon^*_{2\perp}\right)
 -P^\gamma_i(M_{H_i})\frac{2}{M_{H_i}^2}
\langle\epsilon^*_1\epsilon^*_2 k_1k_2\rangle
\bigg\}\,,
\end{eqnarray}
where $k_{1,2}$ are the momenta of the two photons and
$\epsilon_{1,2}$ the wave vectors of the corresponding photons,
$\epsilon^\mu_{1\perp} = \epsilon^\mu_1 - 2k^\mu_1 (k_2 \cdot
\epsilon_1) / M^2_{H_i}$, $\epsilon^\mu_{2\perp} = \epsilon^\mu_2 -
2k^\mu_2 (k_1 \cdot \epsilon_2) / M^2_{H_i}$ and $\langle \epsilon_1
\epsilon_2 k_1 k_2 \rangle \equiv \epsilon_{\mu\nu\rho\sigma}\,
\epsilon_1^\mu \epsilon_2^\nu k_1^\rho k_2^\sigma$. The scalar and
pseudoscalar form factors, retaining only the dominant loop
contributions from the third--generation (s)fermions, $W^\pm$ and
charged Higgs bosons, are given by
\begin{eqnarray}
S^\gamma_i(M_{H_i})&=&2\sum_{f=b,t,\tilde{\chi}^\pm_1,\tilde{\chi}^\pm_2} N_C\, Q_f^2\,
g_fg^{S}_{H_i\bar{f}f}\,\frac{v}{m_f} F_{sf}(\tau_{if}) \nonumber \\
&&
- \sum_{\tilde{f}_j=\tilde{t}_1,\tilde{t}_2,\tilde{b}_1,\tilde{b}_2,
           \tilde{\tau}_1,\tilde{\tau}_2}
N_C\, Q_f^2g_{H_i\tilde{f}^*_j\tilde{f}_j}
\frac{v^2}{2m_{\tilde{f}_j}^2} F_0(\tau_{i\tilde{f}_j})
\nonumber \\
&&- g_{_{H_iVV}}F_1(\tau_{iW})-
g_{_{H_iH^+H^-}}\frac{v^2}{2 M_{H^\pm}^2} F_0(\tau_{iH^\pm})
\,, \nonumber \\
P^\gamma_i(M_{H_i})&=&2\sum_{f=b,t,\tilde{\chi}^\pm_1,\tilde{\chi}^\pm_2}
N_C\,Q_f^2\,g_fg^{P}_{H_i\bar{f}f}
\,\frac{v}{m_f}
F_{pf}(\tau_{if})
 \,,
\end{eqnarray}
where $\tau_{ix}=M_{H_i}^2/4m_x^2$, $N_C=3$ for (s)quarks and $N_C=1$ for
staus and charginos, respectively.

The form factors $F_{sf}$, $F_{pf}$, $F_0$, and $F_1$ can be
expressed in terms of a so-called scaling function $f(\tau)$, by
\begin{eqnarray}
F_{sf}(\tau)&=&\tau^{-1}\,[1+(1-\tau^{-1}) f(\tau)]\,,~~
F_{pf}(\tau)=\tau^{-1}\,f(\tau)\,,\\
F_0(\tau)&=&\tau^{-1}\,[-1+\tau^{-1}f(\tau)]\,, \hspace{1.2 cm}
F_1(\tau)=2+3\tau^{-1}+3\tau^{-1} (2-\tau^{-1} )f(\tau) \,,\nonumber
\label{formfactor}
\end{eqnarray}
where $f(\tau)$ stands for the integrated function
\begin{eqnarray}
f(\tau)=-\frac{1}{2}\int_0^1\frac{{\rm d}y}{y}\ln\left[1-4\tau y(1-y)\right]
       =\left\{\begin{array}{cl}
           {\rm arcsin}^2(\sqrt{\tau}) \,:   & \qquad \tau\leq 1\,, \\
   -\frac{1}{4}\left[\ln \left(\frac{\sqrt{\tau}+\sqrt{\tau-1}}{
                                     \sqrt{\tau}-\sqrt{\tau-1}}\right)
                    -i\pi\right]^2\,: & \qquad \tau\geq 1\,.
\end{array}\right.
\end{eqnarray}
It is clear that imaginary parts of the form factors
appear for Higgs-boson masses greater than twice the mass of
the charged particle running in the loop, i.e., $\tau\geq 1$. In the limit
$\tau\rightarrow 0$, $F_{sf}(0)=2/3$, $F_{pf}(0)=1$, $F_0(0)=1/3$, and
$F_1(0)=7$.
Finally, the decay width is given by
\begin{eqnarray}
\Gamma(H_i\rightarrow \gamma\gamma)=\frac{M_{H_i}^3\alpha^2}{256\pi^3\,v^2}
         \left[\,\left|S^\gamma_i(M_{H_i})\right|^2
              +\left|P^\gamma_i(M_{H_i})\right|^2\right]\,.
\end{eqnarray}
The QCD correction to the width $\Gamma(H_i\rightarrow \gamma\gamma)$ 
is included in the large loop-mass limit
by multiplying the rescaling $J^\gamma_q$ and
$J^\gamma_{\tilde{q}}$ factors 
to the quark and squark contributions to $S^\gamma_i$, respectively.
The rescaling factors in the large loop-mass limit are
given by~\cite{SDGZ}
\begin{equation}
J^\gamma_t=1-\frac{\alpha_s(M_{H_i}^2)}{\pi}\,, \hspace{1 cm}
J^\gamma_{\tilde{q}}=1+\frac{8\alpha_s(M_{H_i}^2)}{3\pi}\,.
\end{equation}
\item \underline{$H_i\rightarrow gg$}:
The amplitude for the decay process
$H_i \rightarrow gg$ ($i=1,2,3$) can be written as
\begin{eqnarray} \label{higg}
{\cal M}_{gg H_i}=-\frac{\alpha_s\,M_{H_i}^2\,\delta^{ab}}{4\pi\,v}
\bigg\{S^g_i(M_{H_i})
\left(\epsilon^*_{1\perp}\cdot\epsilon^*_{2\perp}\right)
 -P^g_i(M_{H_i})\frac{2}{M_{H_i}^2}
\langle\epsilon^*_1\epsilon^*_2 k_1k_2\rangle
\bigg\}\,,
\end{eqnarray}
where $a$ and $b$ ($a,b=1$ to 8) are indices of the eight SU(3)
generators in the adjoint representation, and $k_{1,2}$ and
$\epsilon_{1,2}$ are the four--momenta and wave vectors of the two
gluons, respectively. The scalar and pseudoscalar form factors,
retaining only the dominant contributions from third--generation
(s)quarks, are given by
\begin{eqnarray}
S^g_i(M_{H_i})&=&\sum_{f=b,t}
g_fg^{S}_{H_if\bar{f}}\,\frac{v}{m_f} F_{sf}(\tau_{if})
-\sum_{\tilde{f}_j=\tilde{t}_1,\tilde{t}_2,\tilde{b}_1,\tilde{b}_2}
g_{H_i\tilde{f}^*_j\tilde{f}_j}
\frac{v^2}{4m_{\tilde{f}_j}^2} F_0(\tau_{i\tilde{f}_j}) \,, \nonumber \\
P^g_i(M_{H_i})&=&\sum_{f=b,t}
g_fg^{P}_{H_if\bar{f}}\,\frac{v}{m_f}
F_{pf}(\tau_{if}) \,.
\end{eqnarray}
The decay width of the process $H_i\rightarrow gg$ is then given by
\begin{eqnarray}
\Gamma(H_i\rightarrow gg)\ =\ \frac{M_{H_i}^3\alpha^2_S}{32\pi^3\,v^2}
         \left[\,K^g_{H}\, \left|S^g_i(M_{H_i})\right|^2\:
              +\: K^g_{A}\, \left|P^g_i(M_{H_i})\right|^2\right]\,,
\end{eqnarray}
where $K^g_{H,A}$  are QCD loop  enhancement factors that  include the
leading-order QCD corrections.  In  the heavy-quark limit, the factors
$K^g_{H,A}$ are given by~\cite{SDGZ}\footnote{We ignore the small
difference between the $K-$factors of the quark and squark loop
contributions to $S_i^g$.}
\begin{eqnarray}
  \label{KgHA}
K^g_H &=& 1\ +\ \frac{\alpha_S (M^2_{H_i})}{\pi}\,
\bigg(\,\frac{95}{4} \: -\: \frac{7}{6}\,N_F\,\bigg)\nonumber\\
K^g_A &=& 1\ +\ \frac{\alpha_S (M^2_{H_i})}{\pi}\,
\bigg(\,\frac{97}{4} \: -\: \frac{7}{6}\,N_F\,\bigg)\,,
\end{eqnarray}
where $N_F$ is the number of quark flavours lighter than the $H_i$
boson. Away from isolated regions of the parameter space~\cite{SDGZ}, the
above $K^g_{H,A}$ factors capture the main bulk of the NLO
corrections with an accuracy at the 10\% level.

\end{itemize}

\section{The Structure of {\tt CPsuperH}}\label{sec:code}

The program {\tt CPsuperH} is self--contained with all necessary
subroutines included.  The Fortran code {\tt CPsuperH} uses three
input arrays for reading the input parameters and five output arrays
for generating the Higgs couplings, decay widths and branching
fractions.  The three input arrays are named
\begin{center}
{\tt SMPARA\_H(IP)}, \ \
{\tt SSPARA\_H(IP)}, \ \
{\tt IFLAG\_H(NFLAG)}.
\end{center}
Among the five output arrays, the arrays for generating the Higgs
couplings are named
\begin{center}
{\tt NHC\_H(NC,IH)},  \ \
{\tt SHC\_H(NC)},  \ \
{\tt CHC\_H(NC)},
\end{center}
and the other two output arrays for the decay widths and branching
fractions are named
\begin{center}
{\tt GAMBRN(IM,IWB,IH)},  \ \
{\tt GAMBRC(IM,IWB)}.
\end{center}
The code {\tt CPsuperH} contains also arrays for the masses and mixing
matrices of the Higgs bosons, stops, sbottoms, staus, charginos and
neutralinos as explained below.

\subsection{Input Arrays}

In this subsection, we describe the details of the input arrays.

\begin{itemize}
\item {\tt SMPARA\_H(IP)}: This is the array for the SM input parameters.
In the current version, we are dealing with 15 inputs as shown in Table
\ref{tab:smpara}, but these can easily be extended by changing {\tt NSMIN} 
in
\underline{\tt cpsuperh.f}.  This array is filled from
the file \underline{\tt run}.

\begin{table}[\hbt]
\caption{\label{tab:smpara}
{\it
The contents of {\tt SMPARA\_H(IP)}.  }
}
\begin{center}
\begin{tabular}{|c|c|c|c|c|c|c|c|}
\hline
{\tt IP} & Parameter & {\tt IP} & Parameter
& {\tt IP} & Parameter & {\tt IP} & Parameter \\
\hline
   1 & $\alpha^{-1}_{\rm em}(M_Z)$ &
   6 & $m_\mu$                     &
  11 & $m_u (m_t^{\rm pole})$      &
  16 & ... \\
   2 & $\alpha_s(M_Z)$             &
   7 & $m_\tau $                   &
  12 & $m_c (m_t^{\rm pole})$      &
  17 & ... \\
   3 & $M_Z$                       &
   8 & $m_d (m_t^{\rm pole})$      &
  13 & $m_t^{\rm pole}$            &
  18 & ... \\
   4 & $\sin^2\theta_W$            &
   9 & $m_s (m_t^{\rm pole})$      &
  14 & $\Gamma_W$                  &
  19 & ... \\
   5 & $m_e$                       &
  10 & $m_b (m_t^{\rm pole})$      &
  15 & $\Gamma_Z$                  &
  20 & ... \\
\hline
\end{tabular}
\end{center}
\end{table}
%

\item {\tt SSPARA\_H(IP)}: This array is for the SUSY input parameters.
In the current version, we are dealing with 21 inputs as shown in Table
\ref{tab:sspara}, but these can easily be extended as well by changing 
{\tt
NSSIN} in \underline{\tt cpsuperh.f}. This array is also
filled from the file \underline{\tt run}.

\begin{table}[\hbt]
\caption{\label{tab:sspara}
{\it
The contents of {\tt SSPARA\_H(IP)}.  }
}
\begin{center}
\begin{tabular}{|c|c|c|c|c|c|c|c|c|c|}
\hline
  {\tt IP} & Parameter
& {\tt IP} & Parameter
& {\tt IP} & Parameter
& {\tt IP} & Parameter
& {\tt IP} & Parameter \\
\hline
   1 & $\tan\beta$                  &
   6 & $\Phi_1$                     &
  11 & $m_{\tilde{Q}_3}$            &
  16 & $|A_t|$                      &
  21 & $\Phi_{A_\tau}$ \\
   2 & $M_{H^\pm}^{\rm pole}$       &
   7 & $|M_2|$                      &
  12 & $m_{\tilde{U}_3}$            &
  17 & $\Phi_{A_t}$                 &
  22 & $\ldots$        \\
   3 & $|\mu|$                      &
   8 & $\Phi_2$                     &
  13 & $m_{\tilde{D}_3}$            &
  18 & $|A_b|$                      &
  23 & $\ldots$       \\
   4 & $\Phi_\mu$                   &
   9 & $|M_3|$                      &
  14 & $m_{\tilde{L}_3}$            &
  19 & $\Phi_{A_b}$                 &
  24 & $\ldots$       \\
   5 & $|M_1|$                      &
  10 & $\Phi_3$                     &
  15 & $m_{\tilde{E}_3}$            &
  20 & $A_\tau$                     &
  25 & $\ldots$       \\
\hline
\end{tabular}
\end{center}
\end{table}
%
\item {\tt IFLAG\_H(NFLAG)} : This {\tt NFLAG}--dimensional array
controls {\tt CPsuperH}. This flag array is used for printing options,
calculating options, integer input parameters, error messages, etc.
The default value for every flag is zero. This array also can be
filled from the file \underline{\tt run}. Only a part of {\tt
IFLAG\_H} is being used presently by the code:
\begin{itemize}
\item {\tt IFLAG\_H(1)=1}: Print out the input parameters.
\item {\tt IFLAG\_H(2)=1}: Print out the masses and mixing matrix of
the Higgs bosons.
\item {\tt IFLAG\_H(3)=1}: Print out the masses and mixing matrices of
the stops, sbottoms, tau sneutrino and staus.
\item {\tt IFLAG\_H(4)=1}: Print out the masses and mixing matrices of
the charginos and neutralinos.
\item {\tt IFLAG\_H(5)=IX}: Print out the Higgs-boson couplings.
The couplings of $H_1$, $H_2$, $H_3$ and $H^\pm$ to two particles
will be printed for ${\tt IX} = $1, 2, 3, and 4, respectively,
and the Higgs--boson self-couplings will be printed for {\tt IX}=5.
All these couplings can be printed out altogether by taking
{\tt IX}=6.
\item {\tt IFLAG\_H(6)=IX}: Print out the decay widths and branching
ratios.
The decay widths and branching ratios of $H_1$, $H_2$, $H_3$, and $H^\pm$
will be printed for ${\tt IX} = $1, 2, 3, and 4, respectively.
{\tt IX}=5 is for printing out all the decay widths and branching ratios
of the neutral and charged Higgs bosons.
\item {\tt IFLAG\_H(10)=1}: Do not include the finite threshold corrections to
the top- and bottom-quark Yukawa couplings due to the exchanges of gluinos
and charginos.
\item {\tt IFLAG\_H(11)=1}: Use the effective potential masses for Higgs bosons
instead of their pole masses.
\item {\tt IFLAG\_H(20)= ISMN}. The index {\tt ISMN} is used
for {\tt GAMBRN}, which is an array for the neutral Higgs decay widths,
and its default value is {\tt ISMN} = 50. In general, ({\tt
ISMN}-1) is the maximal number of different decay modes of the neutral
Higgs bosons into SM particles. The value of {\tt ISMN} may be changed in
order to incorporate additional rare decay modes.
The index {\tt ISMN} is reserved for
the subtotal decay width and branching fraction of the decays into the
SM particles in the output array {\tt GAMBRN} (see below).
\item {\tt IFLAG\_H(21)=ISUSYN}. Similarly to
{\tt ISMN}, the index {\tt ISUSYN} is used  for {\tt GAMBRN}, and
its default value is {\tt ISUSYN} = 50, with ({\tt
ISUSYN}-1) being the maximal number of different decay modes of the
neutral Higgs bosons into SUSY particles. The index {\tt ISMN+ISUSYN}
is reserved for the subtotal decay width and branching fraction of the
decays into the SUSY particles, while the index {\tt ISMN+ISUSYN+1} is
used for the total decay width, considering decays into both the SM and
SUSY particles in the output array {\tt GAMBRN} (see below).
\item {\tt IFLAG\_H(22)= ISMC}. The index {\tt ISMC}
is used for {\tt GAMBRC}, which is an array for the charged-Higgs decay
width,
and its default value is {\tt ISMC} = 25.
In general, ({\tt
ISMC}-1) is the maximal number of different decay modes of the charged
Higgs bosons into SM particles.  The index {\tt ISMC} is reserved for
the subtotal decay width and branching fraction of the decays into the
SM particles in the output array {\tt GAMBRC} (see below).
\item {\tt IFLAG\_H(23)=ISUSYC}. Similarly to
{\tt ISMC}, the index {\tt ISUSYC} for {\tt GAMBRC}, with ({\tt
ISUSYC}-1) being the maximal number of different decay modes of the
charged Higgs bosons into SUSY particles.  The index {\tt ISMC+ISUSYC}
is reserved for the subtotal decay width and branching fraction of the
decays into the SUSY particles, while the index {\tt ISMC+ISUSYC+1} is
used for the total decay width considering decays both into the SM and
SUSY particles in the output array {\tt GAMBRC} (see below).
\end{itemize}
In Appendix D, we list all the parameter common blocks filled or calculated
from {\tt SMPARA\_H} and {\tt SSPARA\_H}.
\end{itemize}

\subsection{Output Arrays}
In this subsection, we give detailed descriptions of the output
arrays. Some of the entries of {\tt IFLAG\_H} are reserved for various
error messages. This feature might be helpful when using {\tt
CPsuperH} to scan many parameter points:
\begin{itemize}
\item {\tt IFLAG\_H(50)=1}: This is an error message that appears when
a stop or sbottom squared mass is negative.
\item {\tt IFLAG\_H(51)=1}: This is an error message that appears when
the Higgs--boson mass matrix contains a complex or negative eigenvalue.
\item {\tt IFLAG\_H(52)=1}: This is an error message that appears when
the diagonalization of the Higgs mass matrix is not successful.
\item {\tt IFLAG\_H(53)=1}: This is a warning message that appears when
the second--step improvement in the calculations of the pole masses
is needed.
\item {\tt IFLAG\_H(54)=1}: This is an error message that appears when
the iteration resumming the threshold corrections is not convergent.
\item {\tt IFLAG\_H(55)=1}: This is an error message that appears when
the Yukawa coupling has a non--perturbative value: $|h_t|$ or $|h_b|$ $>$
2.
\item {\tt IFLAG\_H(56)=1}: This is an error message that appears when
a tau sneutrino or a stau squared mass is negative.
\end{itemize}

The main numerical output is stored in the following arrays:
\begin{itemize}
\item {\tt NHC\_H(NC,IH)}: This is an array for the {\tt IH}--th
neutral Higgs boson ($H_{\tt IH}$) couplings to two particles with
index {\tt NC}. Currently, this array is filled up to ${\tt
NC=93}$ as shown in Table~\ref{tab:nhc}.
\begin{table}[\hbt]
\caption{\label{tab:nhc}
{\it
The couplings of the {\tt IH}-th neutral Higgs boson to two particles
specified with the index {\tt NC}, {\tt NHC\_H(NC,IH)}.
For the definitions of the
couplings to two fermions, $g_f$, $g^S$, and $g^P$, see Eq.~(\ref{eq:hff})
and Table~\ref{tab:hff}.
}}
\begin{center}
\begin{tabular}{|cl|cl|cl|cl|}
\hline
{\tt NC} & Coupling & {\tt NC} & Coupling & {\tt NC} &  Coupling &
{\tt NC} & Coupling \\
\hline
{\tt  1} & $g_e$                  &
{\tt 26} & $g_{H_{\tt IH}\bar{t}t}^S$    &
{\tt 51} & $g^P_{H_{\tt IH}\tilde{\chi}^0_2\tilde{\chi}^0_3},$
           $g^P_{H_{\tt IH}\tilde{\chi}^0_3\tilde{\chi}^0_2}$ &
{\tt 76} & $g_{H_{\tt IH}\tilde{b}_1^*\tilde{b}_2}$ \\
{\tt  2} & $g_{H_{\tt IH}e^+e^-}^S$      &
{\tt 27} & $g_{H_{\tt IH}\bar{t}t}^P$    &
{\tt 52} & $g_{\tilde{\chi}^0} $   &
{\tt 77} & $g_{H_{\tt IH}\tilde{b}_2^*\tilde{b}_1}$ \\
{\tt  3} & $g_{H_{\tt IH}e^+e^-}^P$      &
{\tt 28} & $g_{\tilde{\chi}^0} $   &
{\tt 53} & $g^S_{H_{\tt IH}\tilde{\chi}^0_2\tilde{\chi}^0_4},$
           $g^S_{H_{\tt IH}\tilde{\chi}^0_4\tilde{\chi}^0_2}$ &
{\tt 78} & $g_{H_{\tt IH}\tilde{b}_2^*\tilde{b}_2}$ \\
{\tt  4} & $g_\mu$                 &
{\tt 29} & $g^S_{H_{\tt IH}\tilde{\chi}^0_1\tilde{\chi}^0_1}$ &
{\tt 54} & $g^P_{H_{\tt IH}\tilde{\chi}^0_2\tilde{\chi}^0_4},$
           $g^P_{H_{\tt IH}\tilde{\chi}^0_4\tilde{\chi}^0_2}$ &
{\tt 79} & $g_{H_{\tt IH}\tilde{\tau}_1^*\tilde{\tau}_1}$ \\
{\tt  5} & $g_{H_{\tt IH}\mu^+\mu^-}^S$  &
{\tt 30} & $g^P_{H_{\tt IH}\tilde{\chi}^0_1\tilde{\chi}^0_1}$ &
{\tt 55} & $g_{\tilde{\chi}^0} $   &
{\tt 80} & $g_{H_{\tt IH}\tilde{\tau}_1^*\tilde{\tau}_2}$ \\
{\tt  6} & $g_{H_{\tt IH}\mu^+\mu^-}^P$  &
{\tt 31} & $g_{\tilde{\chi}^0} $   &
{\tt 56} & $g^S_{H_{\tt IH}\tilde{\chi}^0_3\tilde{\chi}^0_4},$
           $g^S_{H_{\tt IH}\tilde{\chi}^0_4\tilde{\chi}^0_3}$ &
{\tt 81} & $g_{H_{\tt IH}\tilde{\tau}_2^*\tilde{\tau}_1}$\\
{\tt  7} & $g_\tau$               &
{\tt 32} & $g^S_{H_{\tt IH}\tilde{\chi}^0_2\tilde{\chi}^0_2}$ &
{\tt 57} & $g^S_{H_{\tt IH}\tilde{\chi}^0_3\tilde{\chi}^0_4},$
           $g^S_{H_{\tt IH}\tilde{\chi}^0_4\tilde{\chi}^0_3}$ &
{\tt 82} & $g_{H_{\tt IH}\tilde{\tau}_2^*\tilde{\tau}_2}$\\
{\tt  8} & $g_{H_{\tt IH}\tau^+\tau^-}^S$  &
{\tt 33} & $g^P_{H_{\tt IH}\tilde{\chi}^0_2\tilde{\chi}^0_2}$ &
{\tt 58} & $g_{\tilde{\chi}^\pm} $   &
{\tt 83} & $g_{H_{\tt IH}\tilde{\nu}_\tau^*\tilde{\nu}_\tau}$\\
{\tt  9} & $g_{H_{\tt IH}\tau^+\tau^-}^P$  &
{\tt 34} & $g_{\tilde{\chi}^0} $   &
{\tt 59} & $g^S_{H_{\tt IH}\tilde{\chi}^+_1\tilde{\chi}^-_1}$ &
{\tt 84} & $S_{\tt IH}^g(M_{H_{\tt IH}})$ \\
{\tt 10} & $g_d$                  &
{\tt 35} & $g^S_{H_{\tt IH}\tilde{\chi}^0_3\tilde{\chi}^0_3}$ &
{\tt 60} & $g^P_{H_{\tt IH}\tilde{\chi}^+_1\tilde{\chi}^-_1}$ &
{\tt 85} & $P_{\tt IH}^g(M_{H_{\tt IH}})$ \\
{\tt 11} & $g_{H_{\tt IH}\bar{d}d}^S$    &
{\tt 36} & $g^P_{H_{\tt IH}\tilde{\chi}^0_3\tilde{\chi}^0_3}$ &
{\tt 61} & $g_{\tilde{\chi}^\pm}$ &
{\tt 86} & $g_{_{H_{\tt IH}H^+H^-}}$ \\
{\tt 12} & $g_{H_{\tt IH}\bar{d}d}^P$    &
{\tt 37} & $g_{\tilde{\chi}^0} $   &
{\tt 62} & $g^S_{H_{\tt IH}\tilde{\chi}^+_1\tilde{\chi}^-_2}$ &
{\tt 87} & $g_{_{H_{\tt IH}H^+W^-}}$ \\
{\tt 13} & $g_s$                  &
{\tt 38} & $g^S_{H_{\tt IH}\tilde{\chi}^0_4\tilde{\chi}^0_4}$ &
{\tt 63} & $g^P_{H_{\tt IH}\tilde{\chi}^+_1\tilde{\chi}^-_2}$ &
{\tt 88} & $S_{\tt IH}^\gamma(M_{H_{\tt IH}})$ \\
{\tt 14} & $g_{H_{\tt IH}\bar{s}s}^S$    &
{\tt 39} & $g^P_{H_{\tt IH}\tilde{\chi}^0_4\tilde{\chi}^0_4}$ &
{\tt 64} & $g_{\tilde{\chi}^\pm}$ &
{\tt 89} & $P_{\tt IH}^\gamma(M_{H_{\tt IH}})$ \\
{\tt 15} & $g_{H_{\tt IH}\bar{s}s}^P$    &
{\tt 40} & $g_{\tilde{\chi}^0} $   &
{\tt 65} & $g^S_{H_{\tt IH}\tilde{\chi}^+_2\tilde{\chi}^-_1}$ &
{\tt 90} & $S_{\tt IH}^g(0)$ \\
{\tt 16} & $g_b$                  &
{\tt 41} & $g^S_{H_{\tt IH}\tilde{\chi}^0_1\tilde{\chi}^0_2},$
           $g^S_{H_{\tt IH}\tilde{\chi}^0_2\tilde{\chi}^0_1}$ &
{\tt 66} & $g^P_{H_{\tt IH}\tilde{\chi}^+_2\tilde{\chi}^-_1}$ &
{\tt 91} & $P_{\tt IH}^g(0)$ \\
{\tt 17} & $g_{H_{\tt IH}\bar{b}b}^S$    &
{\tt 42} & $g^P_{H_{\tt IH}\tilde{\chi}^0_1\tilde{\chi}^0_2},$
           $g^P_{H_{\tt IH}\tilde{\chi}^0_2\tilde{\chi}^0_1}$ &
{\tt 67} & $g_{\tilde{\chi}^\pm}$ &
{\tt 92} & $S_{\tt IH}^\gamma(0)$ \\
{\tt 18} & $g_{H_{\tt IH}\bar{b}b}^P$    &
{\tt 43} & $g_{\tilde{\chi}^0} $   &
{\tt 68} & $g^S_{H_{\tt IH}\tilde{\chi}^+_2\tilde{\chi}^-_2}$ &
{\tt 93} & $P_{\tt IH}^\gamma(0)$ \\
{\tt 19} & $g_u$                  &
{\tt 44} & $g^S_{H_{\tt IH}\tilde{\chi}^0_1\tilde{\chi}^0_3},$
           $g^S_{H_{\tt IH}\tilde{\chi}^0_3\tilde{\chi}^0_1}$ &
{\tt 69} & $g^P_{H_{\tt IH}\tilde{\chi}^+_2\tilde{\chi}^-_2}$ &
{\tt 94} & ... \\
{\tt 20} & $g_{H_{\tt IH}\bar{u}u}^S$    &
{\tt 45} & $g^P_{H_{\tt IH}\tilde{\chi}^0_1\tilde{\chi}^0_3},$
           $g^P_{H_{\tt IH}\tilde{\chi}^0_3\tilde{\chi}^0_1}$ &
{\tt 70} & $g_{_{H_{\tt IH}VV}}$ &
{\tt 95} & ... \\
{\tt 21} & $g_{H_{\tt IH}\bar{u}u}^P$    &
{\tt 46} & $g_{\tilde{\chi}^0} $   &
{\tt 71} & $g_{H_{\tt IH}\tilde{t}_1^*\tilde{t}_1}$ &
{\tt 96} & ... \\
{\tt 22} & $g_c$                  &
{\tt 47} & $g^S_{H_{\tt IH}\tilde{\chi}^0_1\tilde{\chi}^0_4},$
           $g^S_{H_{\tt IH}\tilde{\chi}^0_4\tilde{\chi}^0_1}$ &
{\tt 72} & $g_{H_{\tt IH}\tilde{t}_1^*\tilde{t}_2}$ &
{\tt 97} & ... \\
{\tt 23} & $g_{H_{\tt IH}\bar{c}c}^S$    &
{\tt 48} & $g^P_{H_{\tt IH}\tilde{\chi}^0_1\tilde{\chi}^0_4},$
           $g^P_{H_{\tt IH}\tilde{\chi}^0_4\tilde{\chi}^0_1}$ &
{\tt 73} & $g_{H_{\tt IH}\tilde{t}_2^*\tilde{t}_1}$&
{\tt 98} & ... \\
{\tt 24} & $g_{H_{\tt IH}\bar{c}c}^P$    &
{\tt 49} & $g_{\tilde{\chi}^0} $   &
{\tt 74} & $g_{H_{\tt IH}\tilde{t}_2^*\tilde{t}_2}$&
{\tt 99} & ... \\
{\tt 25} & $g_t$                  &
{\tt 50} & $g^S_{H_{\tt IH}\tilde{\chi}^0_2\tilde{\chi}^0_3},$
           $g^S_{H_{\tt IH}\tilde{\chi}^0_3\tilde{\chi}^0_2}$ &
{\tt 75} & $g_{H_{\tt IH}\tilde{b}_1^*\tilde{b}_1}$ &
{\tt 100} & ... \\
\hline
\end{tabular}
\end{center}
\end{table}
\item {\tt SHC\_H(NC)}:
This array is for the self-couplings of Higgs bosons.
Currently, this array is filled up to ${\tt NC=35}$, as shown in
Table~\ref{tab:shc}.
\begin{table}[\hbt]
\caption{\label{tab:shc}
{\it
The trilinear ({\tt NC}=1--13) and quartic ({\tt NC}=14--35)
Higgs--boson self-couplings, {\tt SHC\_H(NC)}.
We note that {\tt SHC\_H(10+IH)=NHC\_H(86,IH)} for {\tt IH}=1-3.
} }
\begin{center}
\begin{tabular}{|cl|cl|cl|cl|}
\hline
{\tt NC} & Coupling & {\tt NC} & Coupling & {\tt NC} &  Coupling &
{\tt NC} & Coupling \\
\hline
{\tt  1} & $g_{_{H_3H_3H_3}}$ &
{\tt 11} & $g_{_{H_1H^+H^-}}$ &
{\tt 21} & $g_{_{H_3H_2H_2H_1}}$ &
{\tt 31} & $g_{_{H_3H_1H^+H^-}}$ \\
{\tt  2} & $g_{_{H_3H_3H_2}}$ &
{\tt 12} & $g_{_{H_2H^+H^-}}$ &
{\tt 22} & $g_{_{H_3H_2H_1H_1}}$ &
{\tt 32} & $g_{_{H_2H_2H^+H^-}}$ \\
{\tt  3} & $g_{_{H_3H_3H_1}}$ &
{\tt 13} & $g_{_{H_3H^+H^-}}$ &
{\tt 23} & $g_{_{H_3H_1H_1H_1}}$ &
{\tt 33} & $g_{_{H_2H_1H^+H^-}}$ \\
{\tt  4} & $g_{_{H_3H_2H_2}}$ &
{\tt 14} & $g_{_{H_3H_3H_3H_3}}$ &
{\tt 24} & $g_{_{H_2H_2H_2H_2}}$ &
{\tt 34} & $g_{_{H_1H_1H^+H^-}}$ \\
{\tt  5} & $g_{_{H_3H_2H_1}}$ &
{\tt 15} & $g_{_{H_3H_3H_3H_2}}$ &
{\tt 25} & $g_{_{H_2H_2H_2H_1}}$ &
{\tt 35} & $g_{_{H^+H^-H^+H^-}}$ \\
{\tt  6} & $g_{_{H_3H_1H_1}}$ &
{\tt 16} & $g_{_{H_3H_3H_3H_1}}$ &
{\tt 26} & $g_{_{H_2H_2H_1H_1}}$ &
{\tt 36} & ... \\
{\tt  7} & $g_{_{H_2H_2H_2}}$ &
{\tt 17} & $g_{_{H_3H_3H_2H_2}}$ &
{\tt 27} & $g_{_{H_2H_1H_1H_1}}$ &
{\tt 37} & ... \\
{\tt  8} & $g_{_{H_2H_2H_1}}$ &
{\tt 18} & $g_{_{H_3H_3H_2H_1}}$ &
{\tt 28} & $g_{_{H_1H_1H_1H_1}}$ &
{\tt 38} & ... \\
{\tt  9} & $g_{_{H_2H_1H_1}}$ &
{\tt 19} & $g_{_{H_3H_3H_1H_1}}$ &
{\tt 29} & $g_{_{H_3H_3H^+H^-}}$ &
{\tt 39} & ... \\
{\tt 10} & $g_{_{H_1H_1H_1}}$ &
{\tt 20} & $g_{_{H_3H_2H_2H_2}}$ &
{\tt 30} & $g_{_{H_3H_2H^+H^-}}$ &
{\tt 40} & ... \\
\hline
\end{tabular}
\end{center}
\end{table}
\item {\tt CHC\_H(NC)}:
This array is for the couplings of the charged Higgs boson to two
particles.
Currently, this array is filled up to ${\tt NC=48}$, as shown in
Table~\ref{tab:chc}.

\begin{table}[\hbt]
\caption{\label{tab:chc}
{\it
The charged Higgs-boson couplings to two particles specified with the
index
{\tt NC}, {\tt CHC\_H(NC)}.
For the definitions of the couplings to two fermions, $g_{ff^\prime}$,
$g^S$, and $g^P$, see Eq.~(\ref{eq:hff}) and Table~\ref{tab:hff}.
}}
\begin{center}
\begin{tabular}{|cl|cl|cl|cl|cl|}
\hline
{\tt NC} & Coupling & {\tt NC} & Coupling & {\tt NC} &  Coupling &
{\tt NC} & Coupling & {\tt NC} & Coupling \\
\hline
{\tt  1} & $g_{\nu_e e}$ &
{\tt 11} & $g^S_{H^+\bar{u} d}$ &
{\tt 21} & $g^P_{H^+\tilde{\chi}^0_1\tilde{\chi}^-_1}$ &
{\tt 31} & $g_{\tilde{\chi}^0\tilde{\chi}^\pm}$ &
{\tt 41} & $g^S_{H^+\tilde{\chi}^0_4\tilde{\chi}^-_2}$ \\
{\tt  2} & $g^S_{H^+\bar{\nu}_e e^-}$ &
{\tt 12} & $g^P_{H^+\bar{u} d}$ &
{\tt 22} & $g_{\tilde{\chi}^0\tilde{\chi}^\pm}$ &
{\tt 32} & $g^S_{H^+\tilde{\chi}^0_3\tilde{\chi}^-_1}$ &
{\tt 42} & $g^P_{H^+\tilde{\chi}^0_4\tilde{\chi}^-_2}$ \\
{\tt  3} & $g^P_{H^+\bar{\nu}_e e^-}$ &
{\tt 13} & $g_{cs}$ &
{\tt 23} & $g^S_{H^+\tilde{\chi}^0_1\tilde{\chi}^-_2}$ &
{\tt 33} & $g^P_{H^+\tilde{\chi}^0_3\tilde{\chi}^-_1}$ &
{\tt 43} & $g_{H^+\tilde{t}^*_1\tilde{b}_1}$ \\
{\tt  4} & $g_{\nu_\mu \mu}$ &
{\tt 14} & $g^S_{H^+\bar{c} s}$ &
{\tt 24} & $g^P_{H^+\tilde{\chi}^0_1\tilde{\chi}^-_2}$ &
{\tt 34} & $g_{\tilde{\chi}^0\tilde{\chi}^\pm}$ &
{\tt 44} & $g_{H^+\tilde{t}^*_1\tilde{b}_2}$ \\
{\tt  5} & $g^S_{H^+\bar{\nu}_\mu \mu^-}$ &
{\tt 15} & $g^P_{H^+\bar{c} s}$ &
{\tt 25} & $g_{\tilde{\chi}^0\tilde{\chi}^\pm}$ &
{\tt 35} & $g^S_{H^+\tilde{\chi}^0_3\tilde{\chi}^-_2}$ &
{\tt 45} & $g_{H^+\tilde{t}^*_2\tilde{b}_1}$ \\
{\tt  6} & $g^P_{H^+\bar{\nu}_\mu \mu^-}$ &
{\tt 16} & $g_{tb}$ &
{\tt 26} & $g^S_{H^+\tilde{\chi}^0_2\tilde{\chi}^-_1}$ &
{\tt 36} & $g^P_{H^+\tilde{\chi}^0_3\tilde{\chi}^-_2}$ &
{\tt 46} & $g_{H^+\tilde{t}^*_2\tilde{b}_2}$ \\
{\tt  7} & $g_{\nu_\tau \tau}$ &
{\tt 17} & $g^S_{H^+\bar{t} b}$ &
{\tt 27} & $g^P_{H^+\tilde{\chi}^0_2\tilde{\chi}^-_1}$ &
{\tt 37} & $g_{\tilde{\chi}^0\tilde{\chi}^\pm}$ &
{\tt 47} & $g_{H^+\tilde{\nu}^*_\tau\tilde{\tau}_1}$ \\
{\tt  8} & $g^S_{H^+\bar{\nu}_\tau \tau^-}$ &
{\tt 18} & $g^P_{H^+\bar{t} b}$ &
{\tt 28} & $g_{\tilde{\chi}^0\tilde{\chi}^\pm}$ &
{\tt 38} & $g^S_{H^+\tilde{\chi}^0_4\tilde{\chi}^-_1}$ &
{\tt 48} & $g_{H^+\tilde{\nu}^*_\tau\tilde{\tau}_2}$ \\
{\tt  9} & $g^P_{H^+\bar{\nu}_\tau \tau^-}$ &
{\tt 19} & $g_{\tilde{\chi}^0\tilde{\chi}^\pm}$ &
{\tt 29} & $g^S_{H^+\tilde{\chi}^0_2\tilde{\chi}^-_2}$ &
{\tt 39} & $g^P_{H^+\tilde{\chi}^0_4\tilde{\chi}^-_1}$ &
{\tt 49} & ... \\
{\tt 10} & $g_{ud}$ &
{\tt 20} & $g^S_{H^+\tilde{\chi}^0_1\tilde{\chi}^-_1}$ &
{\tt 30} & $g^P_{H^+\tilde{\chi}^0_2\tilde{\chi}^-_2}$ &
{\tt 40} & $g_{\tilde{\chi}^0\tilde{\chi}^\pm}$ &
{\tt 50} & ... \\
\hline
\end{tabular}
\end{center}
\end{table}
\item {\tt GAMBRN(IM,IWB,IH)}: This output array is for the decay
width in GeV ({\tt IWB}=1) and branching fraction ({\tt IWB}=2,3) of
the decay mode specified by the index {\tt IM} of the neutral Higgs
bosons $H_{\tt IH}$. The value {\tt IWB}=2 is for the branching
fraction taking into account the decays only into SM particles, and
{\tt IWB}=3 for that taking account both the SM and SUSY decays. By
default, the code takes ${\tt ISMN}={\tt ISUSYN}=50$.
All the decay modes considered are listed in Table~\ref{tab:gambrn}
for specific {\tt IM}. In particular, {\tt
GAMBRN(IM=ISMN+ISUSYN+1,IWB=1,IH)} is the total decay width of the
neutral Higgs boson $H_{\rm IH}$ and {\tt GAMBRN(IM=ISMN,IWB=1,IH)}
and {\tt GAMBRN(IM=ISMN+ISUSYN,IWB=1,IH)} are the subtotal decay
widths into SM particles and into SUSY particles, respectively.
Therefore, we have the following relations for the branching fractions
{\tt IWB}=2,3 :
$${\tt GAMBRN(ISMN,2,IH)}={\tt GAMBRN(ISMN+ISUSYN+1,3,IH)}=1 $$
and
$${\tt GAMBRN(ISMN,3,IH)}\leq 1\,, \ \ {\tt GAMBRN(ISMN+ISUSYN+1,2,IH)}\geq 1.$$
\begin{table}[\hbt]
\caption{\label{tab:gambrn} {\it The decay mode with the index {\tt
IM} of the neutral Higgs bosons $H_{\tt IH}$ used in {\tt
GAMBRN(IM,IWB,IH)}. When {\tt IWB}=1, {\tt IM=ISMN+ISUSYN+1} is for
the total decay width of the neutral Higgs boson $H_{\rm IH}$ and {\tt
IM=ISMN} and {\tt IM=ISMN+ISUSYN} for the subtotal decay widths into
SM particles ({\tt GAMSM}) and into SUSY particles ({\tt GAMSUSY}),
respectively. Note that the current version of {\tt CPsuperH} does not
compute (loop--induced) absorptive phases, i.e., it currently returns
equal decay widths into CP--conjugate final states.} }
\begin{center}
\begin{tabular}{|cl|cl|cl|}
\hline
{\tt IM} & Decay Mode& {\tt IM} & Decay Mode & {\tt IM} &  Decay Mode \\
\hline
{\tt 1}& $H_{\tt IH}\rightarrow e\bar{e}$ & {\tt 11} & $H_{\tt IH} \rightarrow Z Z$
& .. & ...... \\
{\tt 2}& $H_{\tt IH}\rightarrow \mu\bar{\mu}$ & {\tt 12} & $H_{\tt IH} \rightarrow H_1 Z$
& .. & ...... \\
{\tt 3}& $H_{\tt IH}\rightarrow \tau\bar{\tau}$ & {\tt 13} & $H_{\tt IH} \rightarrow H_2 Z$
& .. &  ...... \\
{\tt 4}& $H_{\tt IH}\rightarrow d\bar{d}$ & {\tt 14} & $H_{\tt IH} \rightarrow H_1 H_1$
& .. &  ...... \\
{\tt 5}& $H_{\tt IH}\rightarrow s\bar{s}$ & {\tt 15} & $H_{\tt IH} \rightarrow H_1 H_2$
& .. &  ...... \\
{\tt 6}& $H_{\tt IH}\rightarrow b\bar{b}$ & {\tt 16} & $H_{\tt IH} \rightarrow H_2 H_2$
& .. &  ...... \\
{\tt 7}& $H_{\tt IH}\rightarrow u\bar{u}$ & {\tt 17} & $H_{\tt IH} \rightarrow \gamma\gamma$
& .. &  ...... \\
{\tt 8}& $H_{\tt IH}\rightarrow c\bar{c}$ & {\tt 18} & $H_{\tt IH} \rightarrow g\,g$
& .. &  ...... \\
{\tt 9}& $H_{\tt IH}\rightarrow t\bar{t}$ & .. & ......
& .. &  ...... \\
{\tt 10}& $H_{\tt IH}\rightarrow WW$ & .. & ......
& {\tt ISMN} &  {\tt GAMSM} \\
\hline
{\tt IM} & Decay Mode& {\tt IM} & Decay Mode & {\tt IM} &  Decay Mode \\
\hline
{\tt ISMN+1}&$H_{\tt IH}\rightarrow \widetilde{\chi}^0_1\widetilde{\chi}_1^0$&
{\tt ISMN+11}& $H_{\tt IH}\rightarrow \widetilde{\chi}^+_1\widetilde{\chi}_1^-$
& {\tt ISMN+21} &  $H_{\tt IH}\rightarrow \widetilde{b}^*_2\widetilde{b}_1$\\
{\tt ISMN+2}&$H_{\tt IH}\rightarrow \widetilde{\chi}^0_1\widetilde{\chi}_2^0$&
{\tt ISMN+12} & $H_{\tt IH}\rightarrow \widetilde{\chi}^+_1\widetilde{\chi}_2^-$
& {\tt ISMN+22} &  $H_{\tt IH}\rightarrow \widetilde{b}^*_2\widetilde{b}_2$\\
{\tt ISMN+3}&$H_{\tt IH}\rightarrow \widetilde{\chi}^0_1\widetilde{\chi}_3^0$&
{\tt ISMN+13} & $H_{\tt IH}\rightarrow \widetilde{\chi}^+_2\widetilde{\chi}_1^-$
& {\tt ISMN+23} &  $H_{\tt IH}\rightarrow \widetilde{\tau}^*_1\widetilde{\tau}_1$\\
{\tt ISMN+4}&$H_{\tt IH}\rightarrow \widetilde{\chi}^0_1\widetilde{\chi}_4^0$&
{\tt ISMN+14} & $H_{\tt IH}\rightarrow \widetilde{\chi}^+_2\widetilde{\chi}_2^-$
& {\tt ISMN+24} &  $H_{\tt IH}\rightarrow \widetilde{\tau}^*_1\widetilde{\tau}_2$\\
{\tt ISMN+5}&$H_{\tt IH}\rightarrow \widetilde{\chi}^0_2\widetilde{\chi}_2^0$&
{\tt ISMN+15} & $H_{\tt IH}\rightarrow \widetilde{t}^*_1\widetilde{t}_1$
& {\tt ISMN+25} &  $H_{\tt IH}\rightarrow \widetilde{\tau}^*_2\widetilde{\tau}_1$\\
{\tt ISMN+6}&$H_{\tt IH}\rightarrow \widetilde{\chi}^0_2\widetilde{\chi}_3^0$&
{\tt ISMN+16} & $H_{\tt IH}\rightarrow \widetilde{t}^*_1\widetilde{t}_2$
& {\tt ISMN+26} &  $H_{\tt IH}\rightarrow \widetilde{\tau}^*_2\widetilde{\tau}_2$\\
{\tt ISMN+7}&$H_{\tt IH}\rightarrow \widetilde{\chi}^0_2\widetilde{\chi}_4^0$&
{\tt ISMN+17} & $H_{\tt IH}\rightarrow \widetilde{t}^*_2\widetilde{t}_1$
& {\tt ISMN+27} &  $H_{\tt IH}\rightarrow \widetilde{\nu}^*_\tau\widetilde{\nu}_\tau$\\
{\tt ISMN+8}&$H_{\tt IH}\rightarrow \widetilde{\chi}^0_3\widetilde{\chi}_3^0$&
{\tt ISMN+18} & $H_{\tt IH}\rightarrow \widetilde{t}^*_2\widetilde{t}_2$
& .. &  ...... \\
{\tt ISMN+9}&$H_{\tt IH}\rightarrow \widetilde{\chi}^0_3\widetilde{\chi}_4^0$&
{\tt ISMN+19} & $H_{\tt IH}\rightarrow \widetilde{b}^*_1\widetilde{b}_1$
& {\tt ISMN+ISUSYN} &  {\tt GAMSUSY} \\
{\tt ISMN+10}&$H_{\tt IH}\rightarrow \widetilde{\chi}^0_4\widetilde{\chi}_4^0$&
{\tt ISMN+20} & $H_{\tt IH}\rightarrow \widetilde{b}^*_1\widetilde{b}_2$
& {\tt ISMN+ISUSYN+1} &  {\tt GAMSM+GAMSUSY} \\
\hline
\end{tabular}
\end{center}
\end{table}
%
\item {\tt GAMBRC(IM,IWB)}: This array is for the decay width in GeV ({\tt
IWB}=1)
and branching fraction ({\tt IWB}=2,3) of the decay mode number {\tt IM} of
the charged Higgs boson. The convention
for {\tt IWB} is the same as that for {\tt GAMBRN}.
In the code, ${\tt ISMC}={\tt ISUSYC}=25$ is taken.
The decay modes considered are shown in Table~\ref{tab:gambrc}.
In particular,
{\tt GAMBRC(IM=ISMC+ISUSYC+1,IWB=1)} is the total decay width of
the charged Higgs boson
and {\tt GAMBRC(IM=ISMC,IWB=1)} and {\tt GAMBRC(IM=ISMC+ISUSYC,IWB=1)} are
subtotal decay widths into SM and into SUSY particles, respectively.
Similarly to the case of the neutral Higgs bosons, we have the relations
$${\tt GAMBRC(ISMC,2)}={\tt GAMBRC(ISMC+ISUSYC+1,3)}=1 $$
and
$${\tt GAMBRC(ISMC,3)}\leq 1\,, \ \ {\tt GAMBRC(ISMC+ISUSYC+1,2)}\geq 1.$$
\begin{table}[\hbt]
\caption{\label{tab:gambrc}
{\it
The decay mode with the index {\tt IM}
of the charged Higgs boson used in {\tt GAMBRC(IM,IWB)}.
When {\tt IWB}=1,
{\tt IM=ISMC+ISUSYC+1} is for the total decay width
of the charged Higgs boson and
{\tt IM=ISMC} and {\tt IM=ISMC+ISUSYC} for the subtotal decay
widths into SM particles ({\tt GAMSM})
and into SUSY particles ({\tt GAMSUSY}), respectively.
}}
\begin{center}
\begin{tabular}{|cl|cl|cl|}
\hline
{\tt IM} & Decay Mode \ \ \ \ \ \ \ &
{\tt IM} & Decay Mode \ \ &
{\tt IM} &
Decay Mode \ \\
\hline
1 & $H^+\rightarrow \bar{e}\nu$
  & ..
  & ......
  & {\tt ISMC+9} & $H^+\rightarrow \tilde{t}_1\tilde{b}_1^*$ \\
2 & $H^+\rightarrow \bar{\mu}\nu$
  & {\tt ISMC}
  & {\tt GAMSM}
  & {\tt ISMC+10} & $H^+\rightarrow \tilde{t}_1\tilde{b}_2^*$ \\
3 & $H^+\rightarrow \bar{\tau}\nu$
  & {\tt ISMC+1}
  & $H^+\rightarrow \widetilde{\chi}^0_1\widetilde{\chi}_1^+$
  & {\tt ISMC+11} & $H^+\rightarrow \tilde{t}_2\tilde{b}_1^*$ \\
4 & $H^+\rightarrow u\bar{d}$
  & {\tt ISMC+2}
  & $H^+\rightarrow \widetilde{\chi}^0_2\widetilde{\chi}_1^+$
  & {\tt ISMC+12} & $H^+\rightarrow \tilde{t}_2\tilde{b}_2^*$ \\
5 & $H^+\rightarrow c\bar{s}$
  & {\tt ISMC+3}
  & $H^+\rightarrow \widetilde{\chi}^0_3\widetilde{\chi}_1^+$
  & {\tt ISMC+13} & $H^+\rightarrow \tilde{\nu}_\tau\tilde{\tau}_1^*$ \\
6 & $H^+\rightarrow t\bar{b}$
  & {\tt ISMC+4}
  & $H^+\rightarrow \widetilde{\chi}^0_4\widetilde{\chi}_1^+$
  & {\tt ISMC+14} & $H^+\rightarrow \tilde{\nu}_\tau\tilde{\tau}_2^*$ \\
7 & $H^+\rightarrow H_1W^+$
  & {\tt ISMC+5}
  & $H^+\rightarrow \widetilde{\chi}^0_1\widetilde{\chi}_2^+$
  & .. &  ...... \\
8 & $H^+\rightarrow H_2W^+$
  & {\tt ISMC+6}
  & $H^+\rightarrow \widetilde{\chi}^0_2\widetilde{\chi}_2^+$
  & .. &  ...... \\
.. & ......
   & {\tt ISMC+7}
   & $H^+\rightarrow \widetilde{\chi}^0_3\widetilde{\chi}_2^+$
   & {\tt ISMC+ISUSYC} &  {\tt GAMSUSY} \\
.. & ......
   & {\tt ISMC+8}
   & $H^+\rightarrow \widetilde{\chi}^0_4\widetilde{\chi}_2^+$
   & {\tt ISMC+ISUSYC+1} &  {\tt GAMSM+GAMSUSY} \\
%
\hline
\end{tabular}
\end{center}
\end{table}
\item The code {\tt CPsuperH} contains output arrays for the masses
and mixing matrices of the neutral Higgs bosons, the sfermions, the
charginos, and the neutralinos, named as follows:
\begin{itemize}
\item {\tt {HMASS\_H(3)}}: The masses of the three neutral Higgs bosons,
     $M_{H_i}$.
\item {\tt {OMIX\_H(3,3)}}: The $3\times 3$ Higgs mixing matrix,
     $O_{\alpha i}$.
\item {\tt {STMASS\_H(2)}}: The masses of the stops, $m_{\tilde{t}_i}$.
\item {\tt {STMIX\_H(2,2)}}: The mixing matrix of the stops, $U^{\tilde{t}}_{\alpha i}$.
\item {\tt {SBMASS\_H(2)}}: The masses of the sbottoms, $m_{\tilde{b}_i}$.
\item {\tt {SBMIX\_H(2,2)}}: The mixing matrix of the sbottoms,
      $U^{\tilde{b}}_{\alpha i}$.
\item {\tt {STAUMASS\_H(2)}}: The masses of the staus, $m_{\tilde{\tau}_i}$.
\item {\tt {STAUMIX\_H(2,2)}}: The mixing matrix of the staus,
      $U^{\tilde{\tau}}_{\alpha i}$.
\item {\tt {SNU3MASS\_H}}: The mass of the tau sneutrino, $m_{\tilde{\nu}_\tau}$.
\item {\tt {MC\_H(2)}}: The masses of the charginos,
      $m_{\tilde{\chi}^\pm_i}$.
\item {\tt {UL\_H(2,2)}}: The mixing matrix of the left--handed charginos,
      $(C_L)_{i \alpha}$.
\item {\tt {UR\_H(2,2)}}: The mixing matrix of the right--handed
      charginos, $(C_R)_{i \alpha}$.
\item {\tt {MN\_H(4)}}: The masses of the neutralinos,
      $m_{\tilde{\chi}^0_i}$.
\item {\tt {N\_H(4,4)}}: The mixing matrix of the neutralinos,
      $N_{i \alpha}$.
\end{itemize}
\end{itemize}

\subsection{How to Run {\tt CPsuperH}}

The package {\tt CPsuperH} consists of two text, five Fortran, and
three shell--script files. The main features of the files are as
follows:
\begin{itemize}
\item \underline{Text files}:
\begin{itemize}
\item The file \underline{\tt ARRAY} shows all the arrays described in
the previous two subsections.
\item The file \underline{\tt COMMON} lists the parameter common
blocks, as described in Appendix D.
\end{itemize}
\item \underline{Fortran files}:
\begin{itemize}
\item \underline{\tt cpsuperh.f}
fills all the arrays in \underline{\tt ARRAY} from the shell--script
file \underline{\tt run} by calling the following four Fortran files.
\item \underline{\tt fillpara.f}
fills the common blocks in \underline{\tt COMMON} from {\tt SMPARA\_H}.
and {\tt SSPARA\_H}.
\item \underline{\tt fillhiggs.f}
fills the arrays for the masses and the mixing matrix of the neutral Higgs
bosons, {\tt HMASS\_H} and {\tt OMIX\_H}.
\item \underline{\tt fillcoupl.f}
fills the arrays for the masses and the mixing matrices of the
stops, the sbottoms, the charginos, and the neutralinos as well as
the couplings arrays {\tt NHC\_H}, {\tt SHC\_H}, and {\tt CHC\_H}.
\item \underline{\tt fillgambr.f}
fills the arrays {\tt GAMBRN} and {\tt GAMBRC}.
\end{itemize}
\item \underline{Shell--script files}:
\begin{itemize}
\item \underline{\tt makelib} creates the library file \underline{\tt
libcpsuperh.a} from the four Fortran files of {\tt fillpara.f}, {\tt
fillhiggs.f}, {\tt fillcoupl.f}, and {\tt fillgambr.f}.
\item \underline{\tt compit} creates the execution file \underline{\tt
cpsuperh.exe} by compiling \underline{\tt cpsuperh.f}, linked with the
library \underline{\tt libcpsuperh.a}.
\item \underline{\tt run} supplies {\tt cpsuperh.f} with the input
values for {\tt SMPARA\_H} and {\tt SSPARA\_H} and part of {\tt
IFLAG\_H}, and then shows the results by running \underline{\tt
cpsuperh.exe}.  The example presented in the present work is based on
the so--called CPX scenario with $\Phi_{A_t} = \Phi_{A_b} =
\Phi_{A_\tau} = \Phi_{3} = 90^\circ$, $M_{\rm SUSY}=500$ GeV,
$\tan\beta=5$ and $M_{H^\pm}^{\rm pole}=300$ GeV.  Details may be
found by inspecting the file. We note that, in the example, only ${\tt
IFLAG\_H(1)=1}$ is turned on initially. The user will have to edit
\underline{\tt run} to choose new sets of parameters. The original
version of \underline{\tt run} provides ample explanations of the
various input parameters.
\end{itemize}
\end{itemize}

It is straightforward to run the code {\tt CPsuperH}.
Type `./{\tt makelib}' and `./{\tt compit}' followed by `./{\tt run}':
\begin{center}
\fbox{
{\rm Run {\tt CPsuperH}}: \,\,\,
./{\tt makelib} \,\, $\rightarrow$ \,\,
./{\tt compit} \,\, $\rightarrow$ \,\,
./{\tt run}
}
\end{center}
and then one can see some outputs depending on the values
of ${\tt IFLAG\_H(1-6)}$.

In Appendix E, we show some sample outputs from a {\tt CPsuperH} test
run based on the CPX scenario. All the values for the input parameters
used in the test run $({\tt IFLAG\_H}(1)=1)$, the masses and mixing
matrix of the neutral Higgs bosons $({\tt IFLAG\_H}(2)=1)$, 
the masses and mixing
matrices of the charginos and neutralinos $({\tt IFLAG\_H}(4)=1)$, 
the lightest Higgs boson couplings $({\tt IFLAG\_H}(5)=1)$, and the decay
width and branching fractions of the lightest Higgs boson $({\tt
IFLAG\_H}(6)=1)$ are generated by taking the ${\tt IFLAG\_H}$ values
given in the parentheses.

\begin{figure}[ht]
\hspace{ 0.0cm}
\vspace{-0.5cm}
\centerline{\epsfig{figure=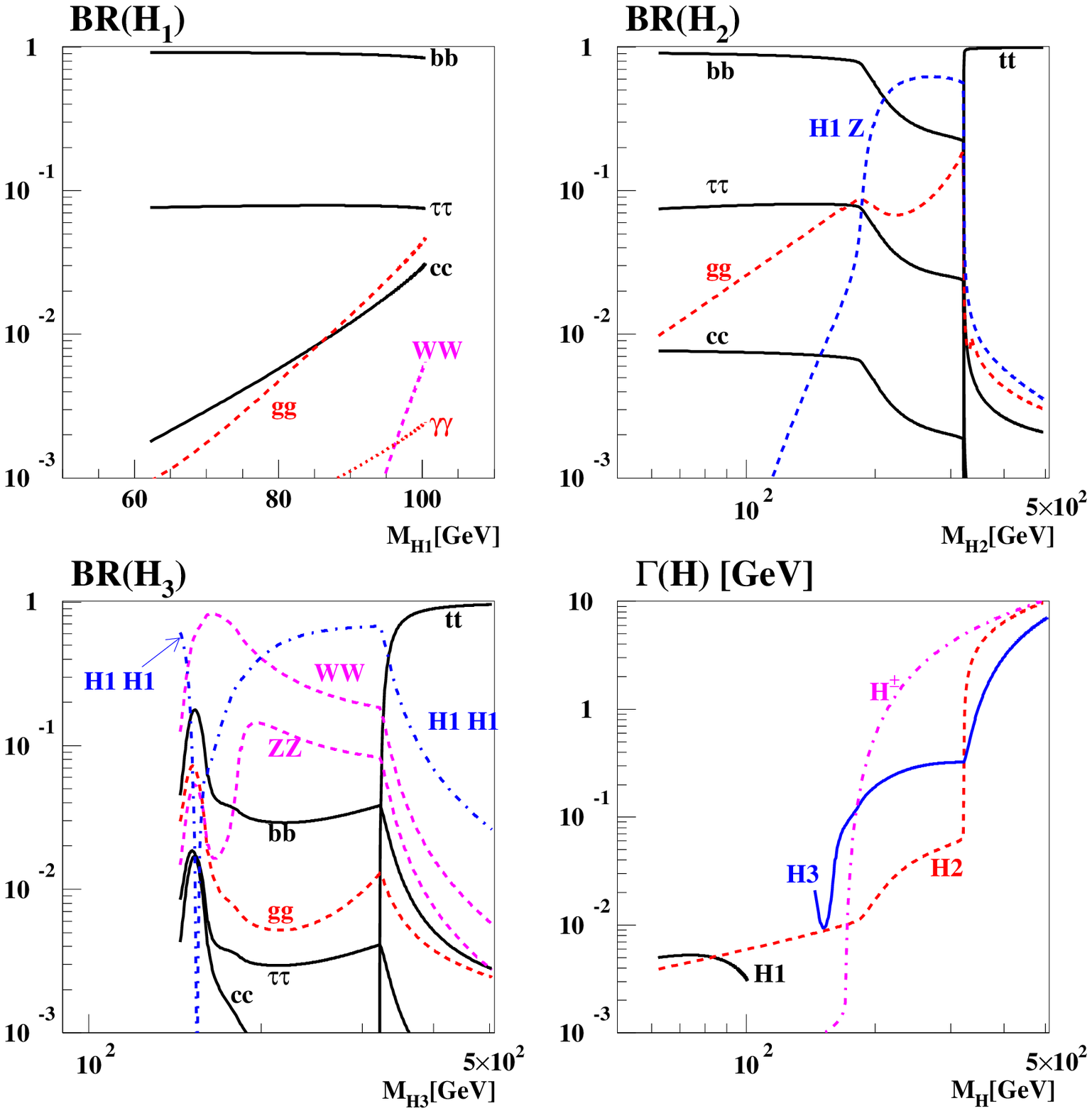,height=19cm,width=19cm}}
\vspace{-0.5cm}
\caption{\it
The branching ratios and total decay widths of the MSSM Higgs bosons,
taking into account only the decays into SM particles. All the CP phases
are set to zero and $\tan\beta=1.5$ is taken.
For the comparison with {\tt HDECAY},  the threshold corrections are
not included and we assume the `maximal mixing' scenario:
$|A_{t,b,\tau}|=\sqrt{6}M_{\rm SUSY}$ with the common SUSY scale
$M_{\rm SUSY}=M_{\tilde{Q}_3}=M_{\tilde{U}_3}=M_{\tilde{D}_3}
=M_{\tilde{L}_3}=M_{\tilde{E}_3}=1$ TeV
and $|\mu|=100$ GeV. The results are consistent with those from the
code {\tt HDECAY}.
}
\label{fig:cpc}
\end{figure}

In order to check whether the code ${\tt CPsuperH}$ generates
numerical outputs consistent with those provided by the code {\tt
HDECAY} \cite{HDECAY} in the CP--invariant case, we run the code ${\tt
CPsuperH}$ in the `maximal mixing' scenario: $|A_{t,b}|=\sqrt{6}M_{\rm
SUSY}$ with the common SUSY scale $M_{\rm
SUSY}=M_{\tilde{Q}_3}=M_{\tilde{U}_3}=M_{\tilde{D}_3}=
M_{\tilde{L}_3}=M_{\tilde{E}_3}=1$ TeV and $|\mu|=100$ GeV, by setting
all the CP phases to zeros, but not including the threshold
corrections. Fig.~\ref{fig:cpc} shows the branching ratios and total
decay widths of the MSSM Higgs bosons as functions of the Higgs boson
masses. Although most of the parameter space presented in
Fig.~\ref{fig:cpc} is already ruled out by Higgs searches at LEP, we
use it to present an effective comparison of the results of the {\tt
CPsuperH} code with those of the {\tt HDECAY} code in the
small-$\tan\beta$ regime.  We find that the {\tt CPsuperH} results are
indeed consistent with those obtained by the code {\tt HDECAY}. There
are very few visible discrepancies, for example in $B(H_1\rightarrow
gg)$ for $M_{H_1}\lsim 80$ GeV, which may be due in part to the
improved calculation of the Higgs-boson mass spectrum and the mixing
matrix in {\tt CPsuperH}.

\begin{figure}[ht]
\hspace{ 0.0cm}
\vspace{-0.5cm}
\centerline{\epsfig{figure=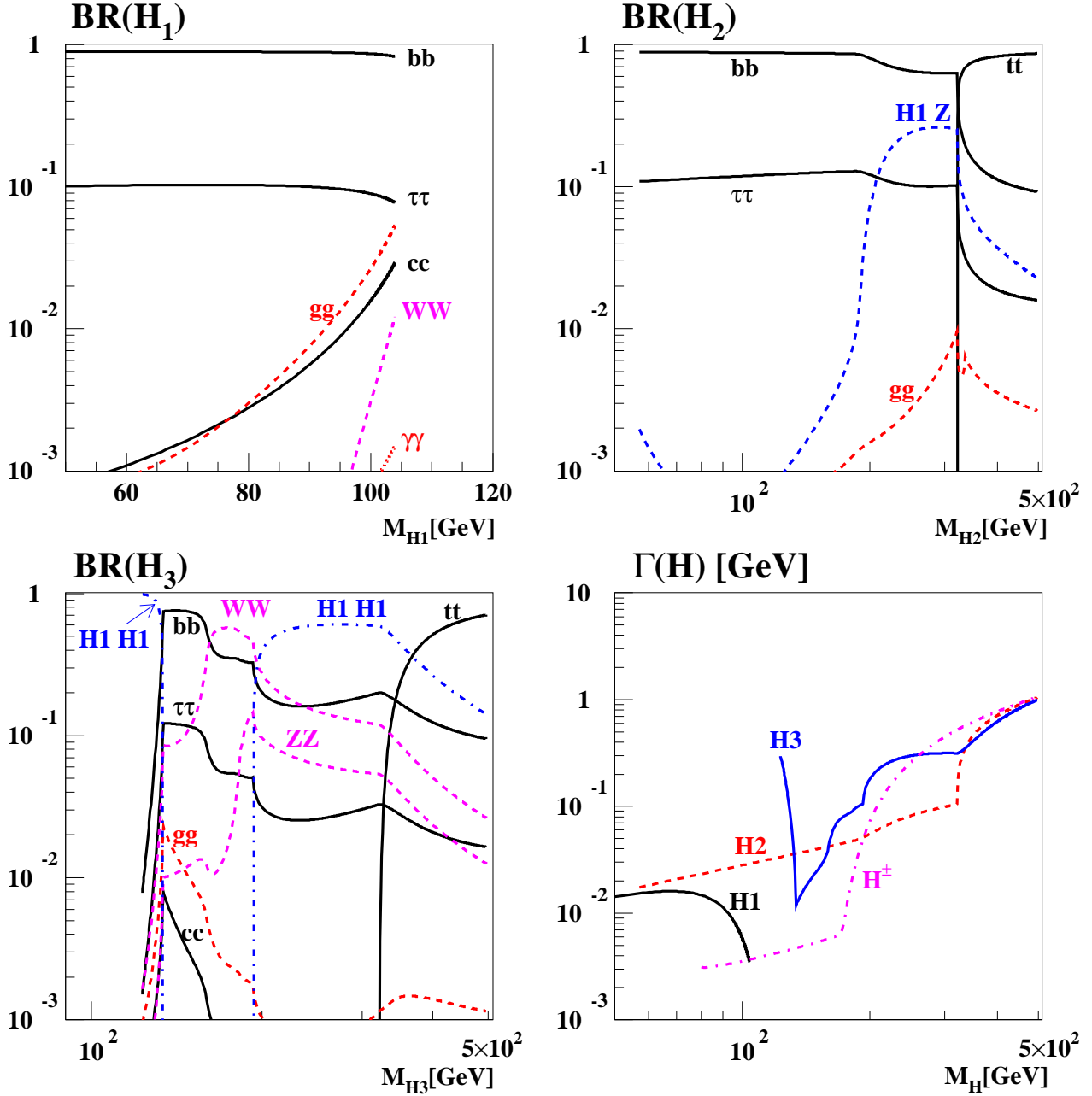,height=19cm,width=19cm}}
\vspace{-0.5cm}
\caption{\it The branching ratios and total decay widths of the MSSM
Higgs bosons, taking into account only the decays into SM
particles. All the CP phases are set to zero and $\tan\beta=4$ is
taken.  We assume the CPX scenario: $|M_3|=1$ TeV, $|\mu|=4M_{\rm
SUSY}$ and $|A_{t,b,\tau}|=2M_{\rm SUSY}$ with the common SUSY scale
$M_{\rm SUSY}=M_{\tilde{Q}_3}=M_{\tilde{U}_3}=M_{\tilde{D}_3}=
M_{\tilde{L}_3}=M_{\tilde{E}_3}=0.5$ TeV.  }
\label{fig:cpx0}
\end{figure}

\begin{figure}[ht]
\hspace{ 0.0cm}
\vspace{-0.5cm}
\centerline{\epsfig{figure=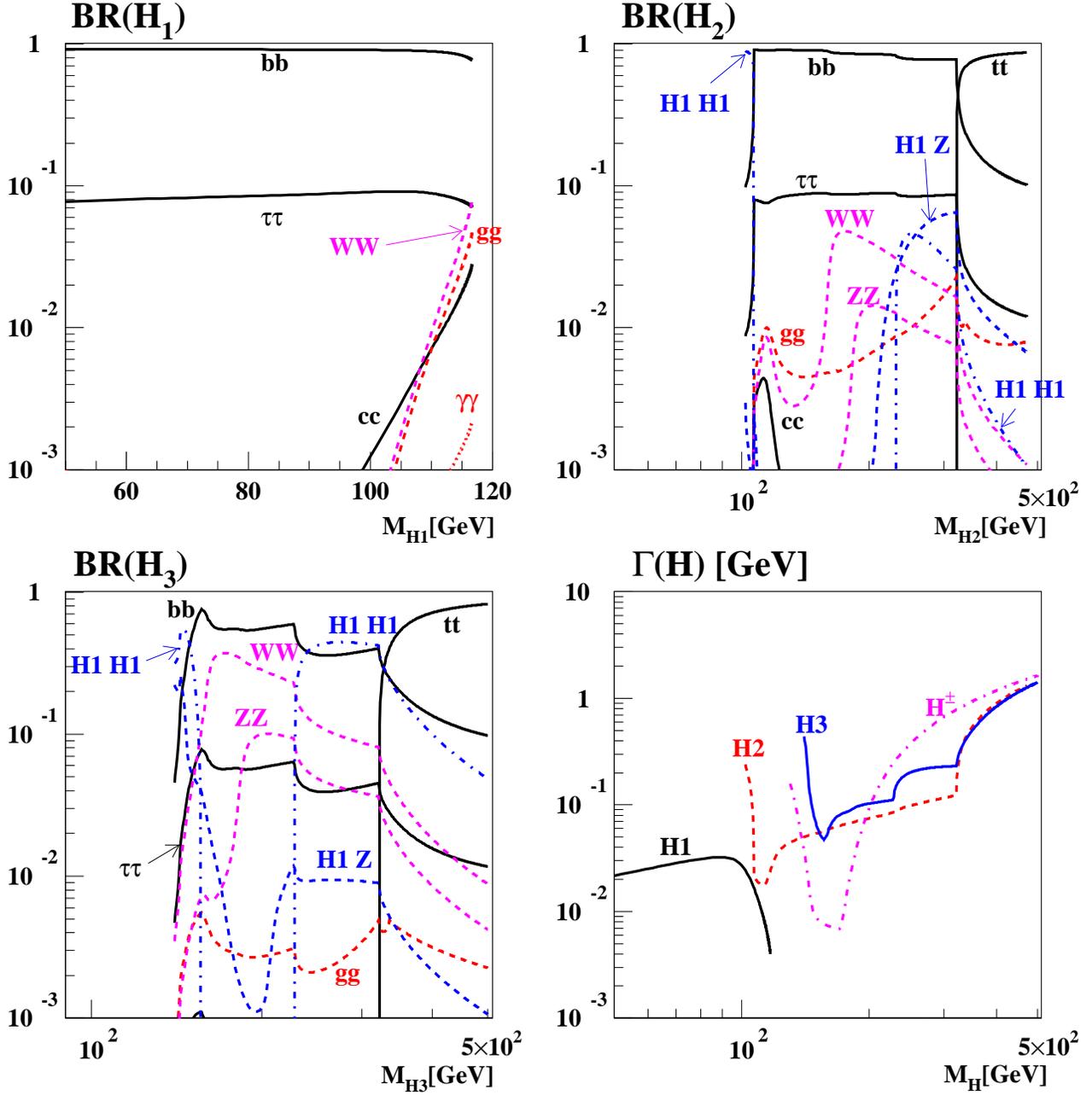,height=19cm,width=19cm}}
\vspace{-0.5cm}
\caption{\it
The same as in Fig.~\ref{fig:cpx0}, but with non--trivial CP phases:
$\Phi_{A_t}=\Phi_{A_b}=\Phi_{A_\tau}=\Phi_3=90^{\rm o}$.
}
\label{fig:cpx1}
\end{figure}

\begin{figure}[ht]
\hspace{ 0.0cm}
\vspace{-0.5cm}
\centerline{\epsfig{figure=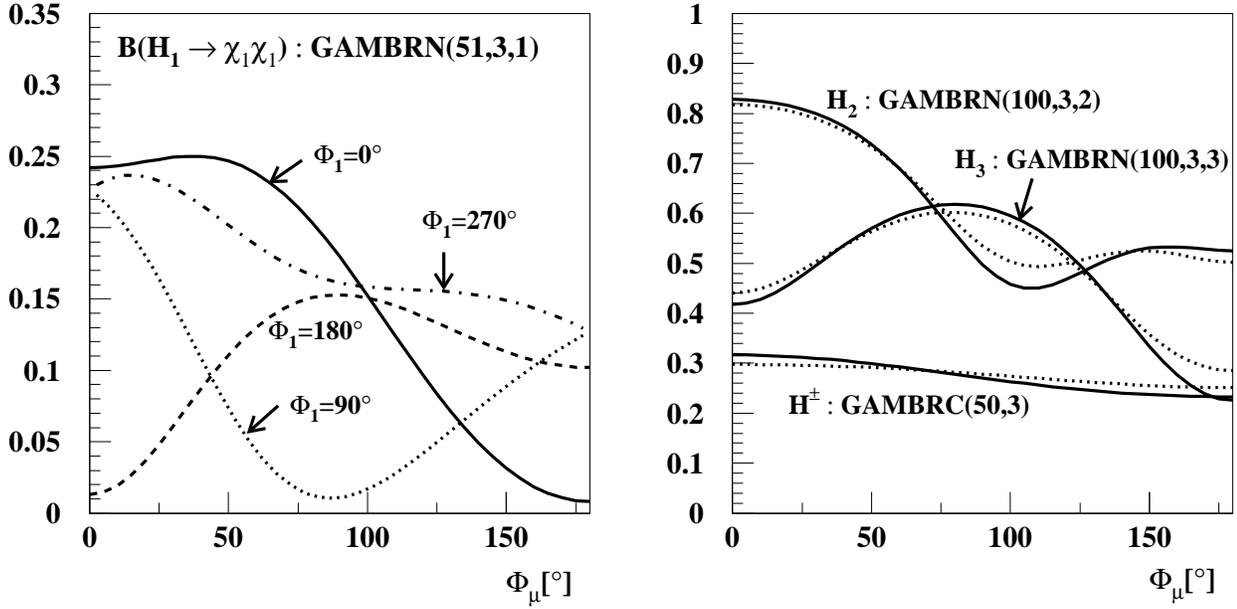,height=19cm,width=19cm}}
\vspace{-8.5cm}
\caption{\it
The dependence of the branching ratios of Higgs bosons into superparticles
on the phase $\phi_\mu$, for $\tan\beta = 5, \, M_{H^+}^{\rm pole} =
0.3$ TeV, $M_{\tilde{Q}_3} = M_{\tilde{U}_3} = M_{\tilde{D}_3} =
M_{\tilde{L}_3} = M_{\tilde{E}_3} = 0.5$ TeV, $|\mu| = 250$ GeV,
$|M_1| = 50$ GeV, $|M_2| = 150$ GeV, $|M_3| = 0.5$ TeV, $|A_t| = |A_b|
= |A_\tau| = 1.2$ TeV, $\Phi_2 = \Phi_3 = 0$, and $\Phi_{A_t} =
\Phi_{A_b} = \Phi_{A_\tau} = - \Phi_\mu$. The left frame shows results
for $H_1$ for several choices of $\Phi_1$; in this case the only
contributing final state is $\tilde\chi_1^0 \tilde \chi_1^0$. The
right frame shows results for the heavier Higgs bosons, where the
solid (dotted) lines are for $\Phi_1 = 0 \ (180^\circ)$; in this case
heavier neutralinos and charginos contribute, but no sfermions.
}
\label{fig:susy}
\end{figure}

We also show in Figs.~\ref{fig:cpx0} and \ref{fig:cpx1} the branching
ratios, {\tt GAMBRN(IM,2,IH)}, and total decay widths, {\tt
GAMBRN(ISMN,1,IH)} and {\tt GAMBRC(ISMC,1,IH)}, found in the CPX
scenario, which are consistent with the results previously reported
in~\cite{CHL}. Finally, in Fig.~\ref{fig:susy} we illustrate the
strong phase dependence of the Higgs boson decay widths into charginos
and neutralinos. The left frame is for the lightest neutral Higgs
boson, which for the given choice of parameters can only decay into
two lightest neutralinos, i.e. {\tt GAMBRN(ISMN+ISUSYN,3,1) =
GAMBRN(ISMN+1,3,1)}. The right frame shows the branching ratios into
superparticles of the heavier neutral Higgs bosons, {\tt
GAMBRN(ISMN+ISUSYN,3,2)} and {\tt GAMBRN(ISMN+ISUSYN,3,3)}, as well as
the charged Higgs boson, {\tt GAMBRC(ISMC+ISUSYC,3)}, which only
receive contributions from chargino/neutralino final states. The
results of this figure are in close agreement with ref.\cite{CHL}.

\section{Summary and Outlook}\label{sec:summary}

We have presented a detailed description of the Fortran code {\tt
CPsuperH}, a new computational package for studying Higgs
phenomenology in the MSSM with explicit CP violation.  Based on recent
RG-improved diagrammatic calculations~\cite{CEPW2}, the program {\tt
CPsuperH} computes the neutral and charged Higgs-boson masses as well
as the $3\times 3$ neutral Higgs-boson mixing matrix $O$ in the
presence of CP violation in the MSSM Higgs sector.  Although the
dominant one- and two-loop contributions to the Higgs-boson
self-energies are incorporated, there are still finite but subdominant
two-loop contributions that may cause shifts of
3--4~GeV~\cite{CHHHWW,Slavich} in the lightest Higgs-boson mass.
These subdominant contributions can be estimated~\cite{3GeV} to be of
comparable size with the dominant three-loop effects.  Because of the
current lack of a detailed three-loop calculation, the subdominant
two--loop contributions are not included in the present version of the
code {\tt CPsuperH}.

In addition to the Higgs mass spectrum, the program {\tt CPsuperH}
computes all the couplings and the decay widths of the neutral Higgs
bosons $H_{1,2,3}$ and the charged Higgs boson $H^+$, incorporating
the most important quantum corrections~\cite{CHL,CPpp}. In particular,
the leading-order QCD corrections are included for the Higgs decays
into photons, gluons and most hadronic channels, so that (with the possible
exception of squark final states) the theoretical uncertainties in
these decay modes are kept small.

The Fortran code {\tt CPsuperH} provides several options that can be
selected in the input files for generating a variety of outputs for
Higgs boson masses, Higgs decays and their respective branching
fractions. Its structure offers the possibility of extending the list
of input data to include flavour textures of off-diagonal trilinear
and/or soft-squark masses, as these arise in some predictive schemes of
soft SUSY breaking, e.g., in minimal supergravity models or in models
with gauge and anomaly mediation. The code {\tt CPsuperH} allows for
straightforward extensions such as the additions of possible lepton-
or quark-flavour-violating decays of the Higgs bosons. Another
possible extension is the inclusion of loop--induced absorptive
phases, which allows to generate CP--odd rate asymmetries in Higgs
boson decays \cite{CPgeneric}.

Radiative Higgs-sector CP violation in the MSSM has a wealth of
implications for many different areas in particle-physics
phenomenology. CP-violating phenomena mediated by Higgs-boson
exchanges may manifest themselves in a number of low-energy
observables such as the electron, neutron and muon electric dipole
moments~\cite{EDMrecent,APEDM,Hdecays,YS}. They may also affect
flavour-changing neutral-current processes and CP asymmetries
involving $K$ and $B$ mesons~\cite{FCNC,resum1,DP}. Moreover,
CP-violating Higgs effects may influence the annihilation rates of
cosmic relics and hence the abundance of dark matter in the
Universe~\cite{DM2}.  Finally, an accurate determination of the Higgs
spectrum in the presence of CP violation is crucial for testing the
viability of electroweak baryogenesis in the MSSM.

{\it To conclude}, the Fortran code {\tt CPsuperH} can be used as a
powerful and efficient computational tool in quantitatively
understanding these various phenomenological subjects, which are
inter-related within the framework of the MSSM with explicit CP
violation. Even in the CP-conserving case, {\tt CPsuperH} is unique in
computing the neutral and charged Higgs-boson couplings and masses
with equally high level of precision, and should be therefore a useful
tool for the study of MSSM Higgs phenomenology at present and future
colliders.

\vspace{-0.2cm}
\subsection*{Acknowledgements}
\vspace{-0.3cm}
\noindent We thank Eri Asakawa, Benedikt Gaissmaier, Kaoru Hagiwara,
Stephen Mrenna and Jeonghyeon Song for fruitful collaborations. The
work of SYC and MD was partially supported by the KOSEF--DFG Joint
Research Project No. 20015-111-02-2. In addition, the work of SYC was
supported in part by the Korea Research Foundation through the grant
KRF--2002--070--C00022 and in part by KOSEF through CHEP at Kyungpook
National University, and the work of MD was supported in part by the
SFB375 of the Deutsche Forschungsgemeinschaft. Finally, the work of
JSL and AP is supported in part by PPARC, under the grant no:
PPA/G/O/2001/00461, and the work of CW is supported in part by the US
DOE, Division of High Energy Physics, under the contract
No. W-31-109-ENG-38.

\newpage


\def\theequation{\Alph{section}.\arabic{equation}}
\begin{appendix}
\setcounter{equation}{0}
\section{Threshold Corrections}

The exchanges of gluinos and charginos
give finite loop--induced threshold corrections to  the Yukawa couplings
$h_{u,d}$, with the structure
\begin{eqnarray}
h_d &=&\frac{\sqrt{2} m_d}{v \cos\beta}\,\frac{1}
{1+(\delta h_d/h_d)+(\Delta h_d/h_d)\tan\beta}\,,
\nonumber \\
h_u &=&\frac{\sqrt{2} m_u}{v \sin\beta}\,\frac{1}
{1+(\delta h_u/h_u)+(\Delta h_u/h_u)\cot\beta}\,,
\end{eqnarray}
modifying the couplings of the neutral Higgs--boson mass eigenstate
$H_i$ to the scalar and pseudoscalar fermion bilinears as follows:
\begin{eqnarray}
  \label{gSHbb}
g^S_{H_i\bar{d}d} & =& {\rm Re}\, \bigg(\,
\frac{1}{1\, +\, \kappa_d\,\tan\beta}\,\bigg)\,
\frac{O_{\phi_1 i}}{\cos\beta}
\ +\ {\rm Re}\, \bigg(\, \frac{\kappa_d}{1\, +\,
\kappa_d\, \tan\beta}\,\bigg)\
\frac{O_{\phi2 i}}{\cos\beta}\nonumber\\
&& +\: {\rm Im}\, \bigg[\,
\frac{ \kappa_d\, (\tan^2\beta\, +\, 1)}{1\, +\,
\kappa_d\, \tan\beta}\,\bigg]\
O_{ai}\, , \nonumber\\[0.35cm]
  \label{gPHbb}
g^P_{H_i\bar{d}d} & =& -\, {\rm Re}\, \bigg(\,
\frac{ \tan\beta\, -\, \kappa_d}{1\, +\, \kappa_d \tan\beta}\,\bigg)\, O_{ai}
\ +\ {\rm Im}\, \bigg(\, \frac{\kappa_d\,\tan\beta}{1\, +\,
\kappa_d\, \tan\beta}\,\bigg)\
\frac{O_{\phi_1 i}}{\cos\beta}\nonumber\\
&&-\: {\rm Im}\, \bigg(\,
\frac{\kappa_d}{1\, +\, \kappa_d\, \tan\beta}\,\bigg)\
\frac{O_{\phi2 i}}{\cos\beta}\ , \\[0.35cm]
  \label{gSHtt}
g^S_{H_i\bar{u}u} & =& {\rm Re}\, \bigg(\,
\frac{1}{1\, +\, \kappa_u\,\cot\beta}\,\bigg)\, \frac{O_{\phi2 i}}{\sin\beta}
\ +\ {\rm Re}\, \bigg(\, \frac{\kappa_u}{1\, +\,
\kappa_u\, \cot\beta}\,\bigg)\
\frac{O_{\phi_1 i}}{\sin\beta}\nonumber\\
&& +\: {\rm Im}\, \bigg[\,
\frac{\kappa_u\, (\cot^2\beta\, +\, 1)}{1\, +\,
\kappa_u\, \cot\beta}\,\bigg]\
O_{ai}\, , \nonumber\\[0.35cm]
  \label{gPHtt}
g^P_{H_i\bar{u}u} & =& -\, {\rm Re}\, \bigg(\,
\frac{\cot\beta\, -\, \kappa_u}{1\, +\,
\kappa_u\,\cot\beta}\,\bigg)\, O_{ai}\ +\
{\rm Im}\, \bigg(\, \frac{\kappa_u\,\cot\beta}{1\, +\,
\kappa_u\, \cot\beta}\,\bigg)\
\frac{O_{\phi_2 i}}{\sin\beta}\nonumber\\
&& -\: {\rm Im}\, \bigg(\,
\frac{\kappa_u}{1\, +\, \kappa_u\, \cot\beta}\,\bigg)\
\frac{O_{\phi_1 i}}{\sin\beta}\ .
\end{eqnarray}
In the above equations, we have used the abbreviation
\begin{equation}
\kappa_q\ =\ \frac{(\Delta h_q/h_q)}{1\: +\: (\delta h_q/h_q)}\, ,
\label{eq:kappa}
\end{equation}
for $q=u,d$. Detailed expressions for the threshold contributions
$(\delta h_q/h_q)$ and $(\Delta h_q/h_q)$ can be found in~\cite{CEMPW}.

For the couplings of the charged Higgs boson to quarks,
we have
\begin{eqnarray}
  \label{gLHud}
g^L_{H^+\bar{u}d} & =& \frac{\cot\beta\, (1\, +\, \rho_u )\:
-\: \bar{\kappa}_u}{1\: +\: \kappa_u\,\cot\beta} \ , \\[0.35cm]
  \label{gRHud}
g^R_{H^+\bar{u}d} & =& \frac{\tan\beta\, (1\, +\, \rho^*_d )\:
-\: \bar{\kappa}^*_d}{1\: +\: \kappa_d^*\,\tan\beta} \ .
\end{eqnarray}
The quantities $\kappa_{u,d}$ are given in (\ref{eq:kappa}),
whilst the quantities $\bar{\kappa}_{u,d}$   and    $\rho_{u,d}$
in (\ref{gLHud})
and~(\ref{gRHud}) are defined as follows:
\begin{equation}
\bar{\kappa}_{q}\ =\ \frac{(\bar{\Delta} h_q/h_q )}{1 + (\delta
h_q/h_q)}\ ,\qquad \rho_q\ =\ \frac{(\bar{\delta} h_q/h_q)\:
-\: (\delta h_q/h_q)}{1 + (\delta h_q/h_q)}\ .
\end{equation}
The expressions for the additional threshold contributions
$(\bar{\delta} h_q/h_q)$ and $(\bar{\Delta} h_q/h_q)$ can also be found
in~\cite{CEMPW}.

\setcounter{equation}{0}
\section{Higgs--Boson Couplings to Third Generation Sfermions}
Here we present the Higgs--sfermion--sfermion couplings in the
weak-interaction
basis. The couplings $\Gamma^{\alpha\tilde{f}^*\tilde{f}}$ are given in the
$(\tilde{f}_L, \tilde{f}_R)$ basis by
\begin{eqnarray}
%
%
\Gamma^{a\tilde{b}^*\tilde{b}} &=& \frac{1}{\sqrt{2}}\left(
\begin{array}{cc}
0 & i\,h_b^*(s_\beta A_b^*+c_\beta \mu) \\
-i\,h_b(s_\beta A_b+c_\beta \mu^*) & 0
\end{array} \right)\,,
\nonumber \\
\Gamma^{\phi_1\tilde{b}^*\tilde{b}} &=& \left(
\begin{array}{cc}
-|h_b|^2vc_\beta+ \frac{1}{4}\left(g^2+\frac{1}{3}g^{\prime 2}\right)vc_\beta&
-\frac{1}{\sqrt{2}}h_b^*A_b^* \\
-\frac{1}{\sqrt{2}}h_bA_b &
-|h_b|^2vc_\beta+ \frac{1}{6}g^{\prime2} vc_\beta
\end{array} \right)\,,
\nonumber \\
\Gamma^{\phi_2\tilde{b}^*\tilde{b}} &=& \left(
\begin{array}{cc}
- \frac{1}{4}\left(g^2+\frac{1}{3}g^{\prime 2}\right)vs_\beta&
\frac{1}{\sqrt{2}}h_b^*\mu \\
\frac{1}{\sqrt{2}}h_b\mu^* &
-\frac{1}{6}g^{\prime2} vs_\beta
\end{array} \right)\,,
\nonumber \\
%
%
\Gamma^{a\tilde{t}^*\tilde{t}} &=& \frac{1}{\sqrt{2}}\left(
\begin{array}{cc}
0 & i\,h_t^*(c_\beta A_t^*+s_\beta \mu) \\
-i\,h_t(c_\beta A_t+s_\beta \mu^*) & 0
\end{array} \right)\,,
\nonumber \\
\Gamma^{\phi_1\tilde{t}^*\tilde{t}} &=& \left(
\begin{array}{cc}
- \frac{1}{4}\left(g^2-\frac{1}{3}g^{\prime 2}\right)vc_\beta&
\frac{1}{\sqrt{2}}h_t^*\mu \\
\frac{1}{\sqrt{2}}h_t\mu^* &
-\frac{1}{3}g^{\prime2} vc_\beta
\end{array} \right)\,,
\nonumber \\
\Gamma^{\phi_2\tilde{t}^*\tilde{t}} &=& \left(
\begin{array}{cc}
-|h_t|^2vs_\beta+ \frac{1}{4}\left(g^2-\frac{1}{3}g^{\prime 2}\right)vs_\beta&
-\frac{1}{\sqrt{2}}h_t^*A_t^* \\
-\frac{1}{\sqrt{2}}h_tA_t &
-|h_t|^2vs_\beta+ \frac{1}{3}g^{\prime2} vs_\beta
\end{array} \right)\,,
\nonumber \\
%
%
\Gamma^{a\tilde{\tau}^*\tilde{\tau}} &=& \frac{1}{\sqrt{2}}\left(
\begin{array}{cc}
0 & i\,h_\tau^*(s_\beta A_\tau^*+c_\beta \mu) \\
-i\,h_\tau(s_\beta A_\tau+c_\beta \mu^*) & 0
\end{array} \right)\,,
\nonumber \\
\Gamma^{\phi_1\tilde{\tau}^*\tilde{\tau}} &=& \left(
\begin{array}{cc}
-|h_\tau|^2vc_\beta+ \frac{1}{4}\left(g^2-g^{\prime 2}\right) vc_\beta&
-\frac{1}{\sqrt{2}}h_\tau^*A_\tau^* \\
-\frac{1}{\sqrt{2}}h_\tau A_\tau &
-|h_\tau|^2vc_\beta+ \frac{1}{2}g^{\prime2} vc_\beta
\end{array} \right)\,,
\nonumber \\
\Gamma^{\phi_2\tilde{\tau}^*\tilde{\tau}} &=& \left(
\begin{array}{cc}
  -\frac{1}{4}\left(g^2- g^{\prime 2}\right) vs_\beta&
\frac{1}{\sqrt{2}}h_\tau^*\mu \\
\frac{1}{\sqrt{2}}h_\tau\mu^* &
-\frac{1}{2}g^{\prime2} vs_\beta
\end{array} \right)\,,
\nonumber\\
\Gamma^{a\tilde{\nu}^*_\tau\tilde{\nu}_\tau} &=& 0\,,
\qquad
\Gamma^{\phi_1\tilde{\nu}^*_\tau\tilde{\nu}_\tau} = -\frac{1}{4}
\left( g^2 + g^{\prime 2} \right) v c_\beta,
\qquad
\Gamma^{\phi_2\tilde{\nu}^*_\tau\tilde{\nu}_\tau} = \frac{1}{4} \left(
g^2 + g^{\prime 2} \right) v s_\beta \, .
\end{eqnarray}
The coupling $\Gamma^{H^+\tilde{u}^*\tilde{d}}$ is given in the LR basis by
\begin{eqnarray}
\Gamma^{H^+\tilde{u}^*\tilde{d}}\ &=&\ \left(
\begin{array}{cc} \frac{1}{\sqrt{2}}\,(|h_u|^2 + |h_d|^2
- g^2)\,v s_\beta c_\beta & h_d^*\, ( s_\beta A^*_d + c_\beta \mu )\\
h_u\,( c_\beta A_u + s_\beta \mu^*) &
\frac{1}{\sqrt{2}}\, h_u h^*_d\, v \end{array}\right)\,
\nonumber\\
\Gamma^{H^+\tilde{\nu}_\tau^*\tilde{\tau}_L}\ &=& \
 \frac{1}{\sqrt{2}}\,(|h_\tau|^2 - g^2)\,v s_\beta\, c_\beta\,,\qquad
\Gamma^{H^+\tilde{\nu}_\tau^*\tilde{\tau}_R}\ = \
  h^*_\tau \left(s_\beta A^*_\tau+c_\beta \mu\right)
\end{eqnarray}

\setcounter{equation}{0}
\section{Higgs--Boson Self--Couplings}

In the following, we list all the effective trilinear and quartic
Higgs--boson self--couplings of the Higgs weak eigenstates, which can
be expressed in terms of the conventional quartic couplings
$\lambda_{1,2,\dots,7}$ of the Higgs potential~\cite{PW} obtained in
an expansion of the effective Higgs potential up to operators of
dimension 4.  The trilinear couplings of the neutral Higgs
bosons~\cite{CHL} are given by
\begin{eqnarray}
g_{_{\phi_1\phi_1\phi_1}} \!&=&\!\! c_\beta\lambda_1\: +\:
\frac{1}{2}\,s_\beta\,\real\lambda_6\,, \nonumber\\
g_{_{\phi_1\phi_1\phi_2}} \!&=&\!\! s_\beta\, \lambda_{34}\:
+\: s_\beta\, \real\lambda_5\: +\: \frac{3}{2}\,
c_\beta\, \real\lambda_6\,,\nonumber\\
g_{_{\phi_1\phi_2\phi_2}} \!&=&\!\! c_\beta\, \lambda_{34}\:
+\: c_\beta\, \real\lambda_5\: +\: \frac{3}{2}\,
s_\beta\, \real\lambda_7\,,\nonumber\\
g_{_{\phi_2\phi_2\phi_2}} \!&=&\!\! s_\beta \lambda_2\: +\:
\frac{1}{2}\, c_\beta\,\real\lambda_7\,,\nonumber\\
g_{_{\phi_1\phi_1 a}} \!&=&\!\! -s_\beta c_\beta\, \imag\lambda_5\: -\:
\frac{1}{2}\, (1 + 2c^2_\beta)\,\imag\lambda_6\,,\nonumber\\
g_{_{\phi_1\phi_2 a}} \!&=&\!\! -2\imag\lambda_5\: -\: s_\beta c_\beta\,
\imag\,(\lambda_6 + \lambda_7)\,,\nonumber\\
g_{_{\phi_2\phi_2 a}} \!&=&\!\! -s_\beta c_\beta\,\imag\lambda_5\:
-\: \frac{1}{2}\, (1 + 2s^2_\beta)\, \imag\lambda_7\,,\nonumber\\
g_{_{\phi_1 aa}} \!&=&\!\! s^2_\beta c_\beta\lambda_1\: +\:
c^3_\beta\,\lambda_{34}\: -\: c_\beta (1 +
s^2_\beta)\, \real\lambda_5\: +\: \frac{1}{2}\, s_\beta (s^2_\beta -
2c^2_\beta)\,\real\lambda_6\: +\:
\frac{1}{2} s_\beta c^2_\beta\, \real\lambda_7\,,\nonumber\\
g_{_{\phi_2 aa}} \!&=&\!\! s_\beta c^2_\beta\lambda_2\: +\:
s^3_\beta\, \lambda_{34}\: -\: s_\beta (1 +
c^2_\beta)\, \real\lambda_5\: +\: \frac{1}{2} s^2_\beta
c_\beta\, \real\lambda_6\: +\: \frac{1}{2}\, c_\beta (c^2_\beta -
2s^2_\beta)\,\real\lambda_7\,,\nonumber\\
g_{_{aaa}} \!&=&\!\! s_\beta c_\beta\, \imag\lambda_5\: -\:
\frac{1}{2}\,s^2_\beta\, \imag\lambda_6\: -\: \frac{1}{2}\,
c^2_\beta\, \imag\lambda_7\, ,
\end{eqnarray}
with $\lambda_{34} = \frac{1}{2}\, (\lambda_3 + \lambda_4)$.
The  effective couplings  $g_{_{\alpha H^+H^-}}$~\cite{CHL} read:
\begin{eqnarray}
g_{_{\phi_1H^+H^-}}  \!\!&=& 2s^2_\beta c_\beta\lambda_1\: +\:
c^3_\beta\lambda_3\: -\: s^2_\beta c_\beta \lambda_4\: -\:
2s^2_\beta c_\beta\, \real \lambda_5\:
+\: s_\beta (s^2_\beta - 2c^2_\beta)\, \real\lambda_6\nonumber\\
&&+\: s_\beta c^2_\beta \real \lambda_7\, ,\nonumber\\
g_{_{\phi_2 H^+H^-}} \!\!&=& 2s_\beta c^2_\beta \lambda_2\: +\:
s^3_\beta\lambda_3\: -\: s_\beta c^2_\beta \lambda_4\: -\:
2s_\beta c^2_\beta \, \real \lambda_5\:
+\: s^2_\beta c_\beta \, \real\lambda_6\nonumber\\
&&+\: c_\beta (c^2_\beta - 2s^2_\beta)\, \real \lambda_7\, ,\nonumber\\
g_{_{aH^+H^-}} \!\!&=& 2s_\beta c_\beta\, \imag\lambda_5\: -\:
s^2_\beta\, \imag\lambda_6\: -\: c^2_\beta\, \imag\lambda_7\, .
\end{eqnarray}
The quartic couplings for the neutral Higgs bosons~\cite{CEMPW} are
\begin{eqnarray}
  \label{4H}
g_{_{\phi_1\phi_1\phi_1\phi_1}} \!&=&\! \frac{1}{4}\, \lambda_1\,,\qquad
g_{_{\phi_1\phi_1\phi_1\phi_2}} \ =\  \frac{1}{2}\, \real\lambda_6\,,\qquad
g_{_{\phi_1\phi_1\phi_2\phi_2}} \ =\
  \frac{1}{2}\,\lambda_{34}\: +\: \frac{1}{2}\,\real\lambda_5\,,\nonumber\\
g_{_{\phi_1\phi_2\phi_2\phi_2}} \!&=&\! \frac{1}{2}\, \real\lambda_7\,,\qquad
g_{_{\phi_2\phi_2\phi_2\phi_2}} \ =\ \frac{1}{4}\,\lambda_2\,,\nonumber\\
g_{_{\phi_1\phi_1\phi_1a}} \!&=&\!
      -\frac{1}{2}\,c_\beta\, \imag\lambda_6\,,\qquad
g_{_{\phi_1\phi_1\phi_2 a}} \ =\
   -c_\beta\,\imag\lambda_5\: -\: \frac{1}{2}\,s_\beta\,
\imag\lambda_6\,,\nonumber\\
g_{_{\phi_1\phi_2\phi_2 a}} \!&=&\!
     -s_\beta\,\imag\lambda_5\: -\: \frac{1}{2}\,c_\beta\,
\imag\lambda_7\,,\qquad
g_{_{\phi_2\phi_2\phi_2 a}}\ =\
-\frac{1}{2}\,s_\beta\,\imag\lambda_7\,,\nonumber\\
g_{_{\phi_1\phi_1 aa}} \!&=&\! \frac{1}{2}\,s^2_\beta\, \lambda_1\: +\:
\frac{1}{2}\, c^2_\beta\, \lambda_{34}\: -\: \frac{1}{2}\, c^2_\beta\,
\real\lambda_5\:
-\: \frac{1}{2}\, s_\beta c_\beta\, \real\lambda_6\,,\nonumber\\
g_{_{\phi_1\phi_2 aa}} \!&=&\! -2s_\beta c_\beta\, \real\lambda_5\: +\:
\frac{1}{2}\, s^2_\beta\, \real\lambda_6\: +\: \frac{1}{2}\,
c^2_\beta\, \real\lambda_7\,,\nonumber\\
g_{_{\phi_2\phi_2 aa}} \!&=&\! \frac{1}{2}\,c^2_\beta\, \lambda_2\: +\:
\frac{1}{2}\, s^2_\beta\, \lambda_{34}\: -\: \frac{1}{2}\, s^2_\beta\,
\real\lambda_5\:
-\: \frac{1}{2}\, s_\beta c_\beta\, \real\lambda_7\,,\nonumber\\
g_{_{\phi_1 aaa}} \!&=&\! s_\beta c^2_\beta\, \imag\lambda_5\: -\:
\frac{1}{2}\, s^2_\beta c_\beta\, \imag\lambda_6\: -\:
\frac{1}{2}\, c^3_\beta\, \imag\lambda_7\,,\nonumber\\
g_{_{\phi_2 aaa}} \!&=&\! s^2_\beta c_\beta\, \imag\lambda_5\: -\:
\frac{1}{2}\, s^3_\beta\, \imag\lambda_6\: -\:
\frac{1}{2}\, s_\beta c^2_\beta\, \imag\lambda_7\,,\nonumber\\
g_{_{aaaa}} \!&=&\! \frac{1}{4}\, g_{_{H^+H^-H^+H^-}}\; ,
\end{eqnarray}
with the quartic coupling of the charged Higgs bosons being given by
\begin{eqnarray}
  \label{4Hplus}
g_{_{H^+H^-H^+H^-}} &=&
s^4_\beta \lambda_1\: +\: c^4_\beta \lambda_2\:
+\: s^2_\beta c^2_\beta (\lambda_3 + \lambda_4)\: +\:
2s^2_\beta c^2_\beta \real\lambda_5\:
-\: 2s^3_\beta c_\beta \real\lambda_6\nonumber\\
&& -\: 2s_\beta c^3_\beta \real\lambda_7\,.
\end{eqnarray}
Finally, the remaining quartic couplings involving the charged Higgs
boson pairs, $g_{\alpha\beta {\scriptscriptstyle H^+H^-}}$, are given
by
\begin{eqnarray}
  \label{2Hplus}
g_{_{\phi_1\phi_1 H^+H^-}} \!&=&\! s^2_\beta\, \lambda_1\: +\:
\frac{1}{2}\, c^2_\beta \, \lambda_3\: -\: s_\beta c_\beta\,
\real\lambda_6\,,\nonumber\\
g_{_{\phi_1\phi_2 H^+H^-}} \!&=&\! -\,s_\beta c_\beta\,\lambda_4\: -\:
2s_\beta c_\beta\, \real\lambda_5\: +\: s^2_\beta\,
\real\lambda_6\: +\: c^2_\beta\, \real\lambda_7\,,\nonumber\\
g_{_{\phi_2\phi_2 H^+H^-}} \!&=&\! c^2_\beta\, \lambda_2\: +\:
\frac{1}{2}\, s^2_\beta \, \lambda_3\: -\: s_\beta c_\beta\,
\real\lambda_7\,,\nonumber\\
g_{_{\phi_1 a H^+H^-}} \!&=&\!
2 s_\beta c^2_\beta\, \imag\lambda_5\: -\: s^2_\beta c_\beta\,
\imag\lambda_6\: -\: c^3_\beta \imag\lambda_7\,,\nonumber\\
g_{_{\phi_2 a H^+H^-}} \!&=&\!
2 s^2_\beta c_\beta\, \imag\lambda_5\: -\: s^3_\beta\,
\imag\lambda_6\: -\: s_\beta c^2_\beta \imag\lambda_7\,,\nonumber\\
g_{_{aa H^+H^-}} \!&=&\! g_{_{H^+H^-H^+H^-}}\; .
\end{eqnarray}

\setcounter{equation}{0}
\section{Common Blocks}

Here we list three common blocks for the SM and SUSY parameters, which are
filled from two input arrays {\tt SMPARA\_H}  and {\tt SSPARA\_H}.
\begin{itemize}
\item {\tt /HC\_SMPARA/}: This common block is for the SM parameters.

{\tt
$~$COMMON /HC\_SMPARA/
AEM\_H,ASMZ\_H,MZ\_H,SW\_H,ME\_H,MMU\_H,MTAU\_H,MDMT\_H\\
.\hspace{39mm},MSMT\_H,MBMT\_H,MUMT\_H,MCMT\_H,MTPOLE\_H,GAMW\_H\\
.\hspace{39mm},GAMZ\_H,EEM\_H,ASMT\_H,CW\_H,TW\_H,MW\_H,GW\_H,GP\_H\\
.\hspace{39mm},V\_H,GF\_H,MTMT\_H\\
}
\begin{eqnarray}
\begin{array}{lll}
{\tt AEM\_H} = \alpha_{\rm em}(M_Z) &
{\tt ASMZ\_H} = \alpha_s(M_Z) &
{\tt MZ\_H} = M_Z \ [{\rm GeV}]\\
{\tt SW\_H} = \sin\theta_W &
{\tt ME\_H} = m_e \ [{\rm GeV}] &
{\tt MMU\_H} = m_\mu \ [{\rm GeV}] \\
{\tt MTAU\_H} = m_\tau \ [{\rm GeV}] &
{\tt MDMT\_H} = m_d(m_t^{\rm pole}) \ [{\rm GeV}] &
{\tt MSMT\_H} = m_s(m_t^{\rm pole}) \ [{\rm GeV}] \\
{\tt MBMT\_H} = m_b(m_t^{\rm pole}) \ [{\rm GeV}] &
{\tt MUMT\_H} = m_u(m_t^{\rm pole}) \ [{\rm GeV}] &
{\tt MCMT\_H} = m_c(m_t^{\rm pole}) \ [{\rm GeV}] \\
{\tt MTPOLE\_H} = m_t^{\rm pole} \ [{\rm GeV}] &
{\tt GAMW\_H} = \Gamma_W \ [{\rm GeV}] &
{\tt GAMZ\_H} = \Gamma_Z \ [{\rm GeV}] \\
{\tt EEM\_H} = e=(4\pi\alpha_{\rm em}(M_Z))^{1/2} &
{\tt ASMT\_H} = \alpha_s(m_t^{\rm pole}) &
{\tt CW\_H} = \cos\theta_W \\
{\tt TW\_H} = \tan\theta_W &
{\tt MW\_H} = M_W = M_Z\cos\theta_W &
{\tt GW\_H} = g=e/\sin\theta_W \\
{\tt GP\_H} = g^\prime=e/\cos\theta_W &
{\tt V\_H} = 2M_W/g &
{\tt GF\_H} = G_F \\
{\tt MTMT\_H} = m_t(m_t^{\rm pole})
\end{array}
\nonumber
\end{eqnarray}
\begin{eqnarray}
\alpha_s(m_t^{\rm pole})&=&\frac{\alpha_s(M_Z)}
{1+\beta_{n_f}\frac{\alpha_s(M_Z)}{4\pi}\ln(m_t^{\rm pole\,2}/M_Z^2) }
\ \ {\rm with}  \ \
\beta_{n_f}=11-\frac{2}{3}n_f\,, \nonumber \\
G_F&=&\frac{\sqrt{2}g^2}{8M_W^2}\,, \nonumber \\
m_t(m_t^{\rm pole})&=&\frac{m_t^{\rm pole} }
{1+\frac{4\alpha_s(m_t^{\rm pole})}{3\pi}}\,.
\end{eqnarray}
\item {\tt /HC\_RSUSYPARA/}: This is for the real SUSY parameters.

{\tt
$~~$COMMON /HC\_RSUSYPARA/
TB\_H,CB\_H,SB\_H,MQ3\_H,MU3\_H,MD3\_H,ML3\_H,ME3\_H \\ 
}
\begin{eqnarray}
\begin{array}{llll}
{\tt TB\_H} = \tan\beta &
{\tt CB\_H} = \cos\beta &
{\tt SB\_H} = \sin\beta &
{\tt MQ3\_H} = M_{\tilde{Q}_3} \\
{\tt MU3\_H} = M_{\tilde{U}_3} &
{\tt MD3\_H} = M_{\tilde{D}_3} &
{\tt ML3\_H} = M_{\tilde{L}_3} &
{\tt ME3\_H} = M_{\tilde{E}_3}
\end{array}
\nonumber
\end{eqnarray}
\item {\tt /HC\_CSUSYPARA/}: This is for the complex SUSY parameters.

{\tt
$~~$COMPLEX*16 MU\_H,M1\_H,M2\_H,M3\_H,AT\_H,AB\_H,ATAU\_H \\
$~~~$COMMON /HC\_CSUSYPARA/ MU\_H,M1\_H,M2\_H,M3\_H,AT\_H,AB\_H,ATAU\_H
}
\begin{eqnarray}
\begin{array}{llll}
{\tt MU\_H} = \mu \ [{\rm GeV}]&
{\tt M1\_H} = M_1 \ [{\rm GeV}]&
{\tt M2\_H} = M_2 \ [{\rm GeV}]&
{\tt M3\_H} = M_3 \ [{\rm GeV}] \\
{\tt AT\_H} = A_t \ [{\rm GeV}]&
{\tt AB\_H} = A_b \ [{\rm GeV}]&
{\tt ATAU\_H} = A_\tau \ [{\rm GeV}] &
{ }
\end{array}
\nonumber
\end{eqnarray}
\end{itemize}

\setcounter{equation}{0}
\section{Sample Outputs}

Here we show the results of a test run of the code {\tt CPsuperH}
for the CPX scenario of MSSM Higgs-sector CP violation.

\begin{itemize}
\item ${\tt IFLAG\_H(1)}=1$: The list of the SM and SUSY input parameters  \\ 
{\tt
 ---------------------------------------------------------\\
 Standard Model Parameters  in /HC\_SMPARA/\\
 ---------------------------------------------------------\\
 AEM\_H~~~~=~0.7812E-02~:~alpha\_em(MZ)                    \\
 ASMZ\_H~~~=~0.1172E+00~:~alpha\_s(MZ)                     \\
 MZ\_H~~~~~=~0.9119E+02~:~Z boson mass in GeV             \\
 SW\_H~~~~~=~0.4808E+00~:~sinTheta\_W                      \\
 ME\_H~~~~~=~0.5000E-03~:~electron mass in GeV            \\
 MMU\_H~~~~=~0.1065E+00~:~muon mass in GeV                \\
 MTAU\_H~~~=~0.1777E+01~:~tau mass in GeV                 \\
 MDMT\_H~~~=~0.6000E-02~:~d-quark mass at M\_t\^{}pole in GeV \\
 MSMT\_H~~~=~0.1150E+00~:~s-quark mass at M\_t\^{}pole in GeV \\
 MBMT\_H~~~=~0.3000E+01~:~b-quark mass at M\_t\^{}pole in GeV \\
 MUMT\_H~~~=~0.3000E-02~:~u-quark mass at M\_t\^{}pole in GeV \\
 MCMT\_H~~~=~0.6200E+00~:~c-quark mass at M\_t\^{}pole in GeV \\
 MTPOLE\_H~=~0.1750E+03~:~t-quark pole mass in GeV        \\
 GAMW\_H~~~=~0.2118E+01~:~Gam\_W in GeV                    \\
 GAMZ\_H~~~=~0.2495E+01~:~Gam\_Z in GeV                    \\
 EEM\_H~~~~=~0.3133E+00~:~e = (4*pi*alpha\_em)\^{}1/2         \\
 ASMT\_H~~~=~0.1072E+00~:~alpha\_s(M\_t\^{}pole)               \\
 CW\_H~~~~~=~0.8768E+00~:~cosTheta\_W                      \\
 TW\_H~~~~~=~0.5483E+00~:~tanTheta\_W                      \\
 MW\_H~~~~~=~0.7996E+02~:~W boson mass MW = MZ*CW         \\
 GW\_H~~~~~=~0.6517E+00~:~SU(2) gauge coupling  gw=e/s\_W  \\
 GP\_H~~~~~=~0.3573E+00~:~U(1)\_Y gauge coupling gp=e/c\_W  \\
 V\_H~~~~~~=~0.2454E+03~:~V = 2 MW / gw                   \\
 GF\_H~~~~~=~0.1174E-04~:~GF=sqrt(2)*gw\^{}2/8 MW\^{}2 in GeV\^{}2\\
 MTMT\_H~~~=~0.1674E+03~:~t-quark mass at M\_t\^{}pole in GeV \\
 ---------------------------------------------------------\\
 Real SUSY Parameters  in /HC\_RSUSYPARA/\\
 ---------------------------------------------------------\\
 TB\_H~~~~~=~0.5000E+01~:~tan(beta)                       \\
 CB\_H~~~~~=~0.1961E+00~:~cos(beta)                       \\
 SB\_H~~~~~=~0.9806E+00~:~sin(beta)                       \\
 MQ3\_H~~~~=~0.5000E+03~:~M\_tilde{Q\_3} in GeV             \\
 MU3\_H~~~~=~0.5000E+03~:~M\_tilde{U\_3} in GeV             \\
 MD3\_H~~~~=~0.5000E+03~:~M\_tilde{D\_3} in GeV             \\
 ML3\_H~~~~=~0.5000E+03~:~M\_tilde{L\_3} in GeV             \\
 ME3\_H~~~~=~0.5000E+03~:~M\_tilde{E\_3} in GeV             \\
 ---------------------------------------------------------\\
 Complex SUSY Parameters  in /HC\_CSUSYPARA/\\
 ---------------------------------------------------------\\
 |MU\_H|~~~~~=~0.2000E+04:Mag. of MU parameter in GeV     \\
 |M1\_H|~~~~~=~0.5000E+02:Mag. of M1 parameter in GeV     \\
 |M2\_H|~~~~~=~0.1000E+03:Mag. of M2 parameter in GeV     \\
 |M3\_H|~~~~~=~0.1000E+04:Mag. of M3 parameter in GeV     \\
 |AT\_H|~~~~~=~0.1000E+04:Mag. of AT parameter in GeV     \\
 |AB\_H|~~~~~=~0.1000E+04:Mag. of AB parameter in GeV     \\
 |ATAU\_H|~~~=~0.1000E+04:Mag. of ATAU parameter in GeV   \\
 ARG(MU\_H)~~=~0.0000E+00:Arg. of MU parameter in Degree  \\
 ARG(M1\_H)~~=~0.0000E+00:Arg. of M1 parameter in Degree  \\
 ARG(M2\_H)~~=~0.0000E+00:Arg. of M2 parameter in Degree  \\
 ARG(M3\_H)~~=~0.9000E+02:Arg. of M3 parameter in Degree  \\
 ARG(AT\_H)~~=~0.9000E+02:Arg. of AT parameter in Degree  \\
 ARG(AB\_H)~~=~0.9000E+02:Arg. of AB parameter in Degree  \\
 ARG(ATAU\_H)=~0.9000E+02:Arg. of ATAU parameter in Degree\\
 ---------------------------------------------------------\\
 Charged Higgs boson pole mass : 0.3000E+03 GeV          \\
 ---------------------------------------------------------\\
}
\item ${\tt IFLAG\_H(2)}=1$: The masses and mixing matrix of the
neutral Higgs bosons \\ {\tt
 ---------------------------------------------------------\\
  Masses and Mixing Matrix of Higgs bosons :\\
 $~~~~~~~~~~~~~~~~~~~~~~~~~~~~~~~$HMASS\_H(I) and OMIX\_H(A,I)\\
 ---------------------------------------------------------\\
  H1~~Pole~Mass~~~~~~~~~~~=~0.1188E+03 GeV\\
  H2~~Pole~Mass~~~~~~~~~~~=~0.2703E+03 GeV\\
  H3~~Pole~Mass~~~~~~~~~~~=~0.2981E+03 GeV\\
  Charged~Higgs~Pole~Mass~=~0.3000E+03 GeV [SSPARA\_H(2)]\\
 $~~~~~~~~~~~~~~~~~~~~~~~~~$[H1]~~~~~~~[H2]~~~~~~~[H3]\\
 $~~~~~~~~~~$[phi\_1] / 0.2451E+00  -.3373E+00  -.9089E+00  $\backslash$\ \\
  O(IA,IH)=[phi\_2] | 0.9694E+00  0.7532E-01  0.2335E+00  |\\
 $~~~~~~~~~~$ \ [  a  ] $\backslash$ -.1030E-01  -.9384E+00  0.3454E+00  /\\
 ---------------------------------------------------------\\
}
\item ${\tt IFLAG\_H(4)}=1$: The masses and mixing matrices of the
charginos and neutralinos \\ {\tt
 ---------------------------------------------------------\\
  Chargino Masses and Mixing Matrices :\\
  $~~~~~~~~~~~~~~~~~~~~~~~~$MC\_H(I), UL\_H(I,A), and UR\_H(I,A)\\
 ---------------------------------------------------------\\
  MC1 = 0.9861E+02 GeV~~~~~~MC2 = 0.2003E+04 GeV\\
\\
  UL\_H =\\
   /(0.9984E+00 0.0000E+00) (-.5603E-01 0.0000E+00) $\backslash$\\
   $\backslash$(0.5603E-01 0.0000E+00) (0.9984E+00 0.0000E+00) /\\
\\
  UR\_H =\\
   /(0.9999E+00 0.0000E+00) (-.1385E-01 0.5460E-09) $\backslash$\\
   $\backslash$(0.1385E-01 0.0000E+00) (0.9999E+00 0.0000E+00) /\\
 ---------------------------------------------------------\\
  Neutralino Masses MN\_H(I) and Mixing Matrix N\_H(I,A)\\
 ---------------------------------------------------------\\
  MN1 = 0.4960E+02 GeV~~~~~~MN2 = 0.9862E+02 GeV\\
  MN3 = 0.2001E+04 GeV~~~~~~MN4 = 0.2003E+04 GeV\\
\\
  N\_H(1,1) = (0.9996E+00 0.0000E+00)\\
  N\_H(1,2) = (-.1462E-01 0.0000E+00)\\
  N\_H(1,3) = (0.2218E-01 0.0000E+00)\\
  N\_H(1,4) = (-.4962E-02 0.0000E+00)\\
\\
  N\_H(2,1) = (-.1553E-01 0.0000E+00)\\
  N\_H(2,2) = (-.9991E+00 0.0000E+00)\\
  N\_H(2,3) = (0.3931E-01 0.0000E+00)\\
  N\_H(2,4) = (-.9704E-02 0.0000E+00)\\
\\
  N\_H(3,1) = (0.0000E+00 -.1186E-01)\\
  N\_H(3,2) = (0.0000E+00 0.2112E-01)\\
  N\_H(3,3) = (0.0000E+00 0.7066E+00)\\
  N\_H(3,4) = (0.0000E+00 0.7072E+00)\\
\\
  N\_H(4,1) = (0.1867E-01 0.0000E+00)\\
  N\_H(4,2) = (-.3494E-01 0.0000E+00)\\
  N\_H(4,3) = (-.7062E+00 0.0000E+00)\\
  N\_H(4,4) = (0.7069E+00 0.0000E+00)\\
 ---------------------------------------------------------\\
}
\item ${\tt IFLAG\_H(5)}=1$: The couplings of the lightest Higgs boson \\
{\tt
 ---------------------------------------------------------\\
 The~Lightest~Higgs~H\_1~Couplings~:~NHC\_H(NC,1)\\
 ---------------------------------------------------------\\
 H1 e e~~~~~~~~~~~~~~~[NC=~1]:~GF=(0.2038E-05,0.0000E+00)\\
 $~~~~~~~~~~~~~~~~~~~~~~$[NC=~2]:~GS=(0.1250E+01,0.0000E+00)\\
 $~~~~~~~~~~~~~~~~~~~~~~$[NC=~3]:~GP=(0.5148E-01,0.0000E+00)\\
 H1 mu mu~~~~~~~~~~~~~[NC=~4]:~GF=(0.4340E-03,0.0000E+00)\\
 $~~~~~~~~~~~~~~~~~~~~~~$[NC=~5]:~GS=(0.1250E+01,0.0000E+00)\\
 $~~~~~~~~~~~~~~~~~~~~~~$[NC=~6]:~GP=(0.5148E-01,0.0000E+00)\\
 H1 tau tau~~~~~~~~~~~[NC=~7]:~GF=(0.7242E-02,0.0000E+00)\\
 $~~~~~~~~~~~~~~~~~~~~~~$[NC=~8]:~GS=(0.1250E+01,0.0000E+00)\\
 $~~~~~~~~~~~~~~~~~~~~~~$[NC=~9]:~GP=(0.5148E-01,0.0000E+00)\\
 H1 d d~~~~~~~~~~~~~~~[NC=10]:~GF=(0.2445E-04,0.0000E+00)\\
 $~~~~~~~~~~~~~~~~~~~~~~$[NC=11]:~GS=(0.1250E+01,0.0000E+00)\\
 $~~~~~~~~~~~~~~~~~~~~~~$[NC=12]:~GP=(0.5148E-01,0.0000E+00)\\
 H1 s s~~~~~~~~~~~~~~~[NC=13]:~GF=(0.4687E-03,0.0000E+00)\\
 $~~~~~~~~~~~~~~~~~~~~~~$[NC=14]:~GS=(0.1250E+01,0.0000E+00)\\
 $~~~~~~~~~~~~~~~~~~~~~~$[NC=15]:~GP=(0.5148E-01,0.0000E+00)\\
 H1 b b~~~~~~~~~~~~~~~[NC=16]:~GF=(0.1223E-01,0.0000E+00)\\
 $~~~~~~~~~~~~~~~~~~~~~~$[NC=17]:~GS=(0.1246E+01,0.0000E+00)\\
 $~~~~~~~~~~~~~~~~~~~~~~$[NC=18]:~GP=(-.1741E-01,0.0000E+00)\\
 H1 u u~~~~~~~~~~~~~~~[NC=19]:~GF=(0.1223E-04,0.0000E+00)\\
 $~~~~~~~~~~~~~~~~~~~~~~$[NC=20]:~GS=(0.9886E+00,0.0000E+00)\\
 $~~~~~~~~~~~~~~~~~~~~~~$[NC=21]:~GP=(0.2059E-02,0.0000E+00)\\
 H1 c c~~~~~~~~~~~~~~~[NC=22]:~GF=(0.2527E-02,0.0000E+00)\\
 $~~~~~~~~~~~~~~~~~~~~~~$[NC=23]:~GS=(0.9886E+00,0.0000E+00)\\
 $~~~~~~~~~~~~~~~~~~~~~~$[NC=24]:~GP=(0.2059E-02,0.0000E+00)\\
 H1 t t~~~~~~~~~~~~~~~[NC=25]:~GF=(0.6821E+00,0.0000E+00)\\
 $~~~~~~~~~~~~~~~~~~~~~~$[NC=26]:~GS=(0.9892E+00,0.0000E+00)\\
 $~~~~~~~~~~~~~~~~~~~~~~$[NC=27]:~GP=(0.4501E-02,0.0000E+00)\\
 H1 N1 N1~~~~~~~~~~~~~[NC=28]:~GF=(0.3258E+00,0.0000E+00)\\
 $~~~~~~~~~~~~~~~~~~~~~~$[NC=29]:~GS=(-.5767E-02,0.0000E+00)\\
 $~~~~~~~~~~~~~~~~~~~~~~$[NC=30]:~GP=(0.1317E-03,0.0000E+00)\\
 H1 N2 N2~~~~~~~~~~~~~[NC=31]:~GF=(0.3258E+00,0.0000E+00)\\
 $~~~~~~~~~~~~~~~~~~~~~~$[NC=32]:~GS=(-.1886E-01,0.0000E+00)\\
 $~~~~~~~~~~~~~~~~~~~~~~$[NC=33]:~GP=(0.4125E-03,0.0000E+00)\\
 H1 N3 N3~~~~~~~~~~~~~[NC=34]:~GF=(0.3258E+00,0.0000E+00)\\
 $~~~~~~~~~~~~~~~~~~~~~~$[NC=35]:~GS=(0.1415E-01,0.0000E+00)\\
 $~~~~~~~~~~~~~~~~~~~~~~$[NC=36]:~GP=(0.1576E-03,0.0000E+00)\\
 H1 N4 N4~~~~~~~~~~~~~[NC=37]:~GF=(0.3258E+00,0.0000E+00)\\
 $~~~~~~~~~~~~~~~~~~~~~~$[NC=38]:~GS=(0.3878E-01,0.0000E+00)\\
 $~~~~~~~~~~~~~~~~~~~~~~$[NC=39]:~GP=(-.3866E-03,0.0000E+00)\\
 H1 N1 N2~~~~~~~~~~~~~[NC=40]:~GF=(0.3258E+00,0.0000E+00)\\
 $~~~~~~~~~~~~~~~~~~~~~~$[NC=41]:~GS=(-.1043E-01,0.0000E+00)\\
 $~~~~~~~~~~~~~~~~~~~~~~$[NC=42]:~GP=(0.2331E-03,0.0000E+00)\\
 H1 N1 N3~~~~~~~~~~~~~[NC=43]:~GF=(0.3258E+00,0.0000E+00)\\
 $~~~~~~~~~~~~~~~~~~~~~~$[NC=44]:~GS=(-.1602E-02,0.0000E+00)\\
 $~~~~~~~~~~~~~~~~~~~~~~$[NC=45]:~GP=(0.1443E+00,0.0000E+00)\\
 H1 N1 N4~~~~~~~~~~~~~[NC=46]:~GF=(0.3258E+00,0.0000E+00)\\
 $~~~~~~~~~~~~~~~~~~~~~~$[NC=47]:~GS=(0.2413E+00,0.0000E+00)\\
 $~~~~~~~~~~~~~~~~~~~~~~$[NC=48]:~GP=(-.2402E-02,0.0000E+00)\\
 H1 N2 N3~~~~~~~~~~~~~[NC=49]:~GF=(0.3258E+00,0.0000E+00)\\
 $~~~~~~~~~~~~~~~~~~~~~~$[NC=50]:~GS=(-.2820E-02,0.0000E+00)\\
 $~~~~~~~~~~~~~~~~~~~~~~$[NC=51]:~GP=(0.2540E+00,0.0000E+00)\\
 H1 N2 N4~~~~~~~~~~~~~[NC=52]:~GF=(0.3258E+00,0.0000E+00)\\
 $~~~~~~~~~~~~~~~~~~~~~~$[NC=53]:~GS=(0.4247E+00,0.0000E+00)\\
 $~~~~~~~~~~~~~~~~~~~~~~$[NC=54]:~GP=(-.4228E-02,0.0000E+00)\\
 H1 N3 N4~~~~~~~~~~~~~[NC=55]:~GF=(0.3258E+00,0.0000E+00)\\
 $~~~~~~~~~~~~~~~~~~~~~~$[NC=56]:~GS=(-.2470E-03,0.0000E+00)\\
 $~~~~~~~~~~~~~~~~~~~~~~$[NC=57]:~GP=(-.2789E-03,0.0000E+00)\\
 H1 C1+ C1-~~~~~~~~~~~[NC=58]:~GF=(0.4608E+00,0.0000E+00)\\
 $~~~~~~~~~~~~~~~~~~~~~~$[NC=59]:~GS=(-.2714E-01,0.0000E+00)\\
 $~~~~~~~~~~~~~~~~~~~~~~$[NC=60]:~GP=(0.5936E-03,-.2135E-18)\\
 H1 C1+ C2-~~~~~~~~~~~[NC=61]:~GF=(0.4608E+00,0.0000E+00)\\
 $~~~~~~~~~~~~~~~~~~~~~~$[NC=62]:~GS=(0.6058E+00,0.4035E-02)\\
 $~~~~~~~~~~~~~~~~~~~~~~$[NC=63]:~GP=(-.6042E-02,-.3618E+00)\\
 H1 C2+ C1-~~~~~~~~~~~[NC=64]:~GF=(0.4608E+00,0.0000E+00)\\
 $~~~~~~~~~~~~~~~~~~~~~~$[NC=65]:~GS=(0.6058E+00,-.4035E-02)\\
 $~~~~~~~~~~~~~~~~~~~~~~$[NC=66]:~GP=(-.6042E-02,0.3618E+00)\\
 H1 C2+ C2-~~~~~~~~~~~[NC=67]:~GF=(0.4608E+00,0.0000E+00)\\
 $~~~~~~~~~~~~~~~~~~~~~~$[NC=68]:~GS=(0.5771E-01,0.0000E+00)\\
 $~~~~~~~~~~~~~~~~~~~~~~$[NC=69]:~GP=(-.2527E-03,0.7454E-19)\\
 H1 V V~~~~~~~~~~~~~~~[NC=70]:~G =(0.9987E+00,0.0000E+00)\\
 H1 ST1* ST1~~~~~~~~~~[NC=71]:~G =(0.2184E+01,-.3755E-16)\\
 H1 ST1* ST2~~~~~~~~~~[NC=72]:~G =(0.2447E+00,-.1144E+00)\\
 H1 ST2* ST1~~~~~~~~~~[NC=73]:~G =(0.2447E+00,0.1144E+00)\\
 H1 ST2* ST2~~~~~~~~~~[NC=74]:~G =(-.3987E+01,-.4869E-17)\\
 H1 SB1* SB1~~~~~~~~~~[NC=75]:~G =(0.4330E+00,0.1506E-18)\\
 H1 SB1* SB2~~~~~~~~~~[NC=76]:~G =(0.2070E-01,0.8020E-02)\\
 H1 SB2* SB1~~~~~~~~~~[NC=77]:~G =(0.2070E-01,-.8020E-02)\\
 H1 SB2* SB2~~~~~~~~~~[NC=78]:~G =(-.5584E+00,0.1683E-18)\\
 H1 STA1* STA1~~~~~~~~[NC=79]:~G =(0.2300E+00,-.1144E-17)\\
 H1 STA1* STA2~~~~~~~~[NC=80]:~G =(0.1602E-02,0.6836E-02)\\
 H1 STA2* STA1~~~~~~~~[NC=81]:~G =(0.1602E-02,-.6836E-02)\\
 H1 STA2* STA2~~~~~~~~[NC=82]:~G =(-.3549E+00,-.1785E-17)\\
 H1 SNU3* SNU3~~~~~~~~[NC=83]:~G =(0.1246E+00,0.0000E+00)\\
 H1 glue glue~~~~~~~~~[NC=84]:~S =(0.5827E+00,0.3665E-01)\\
 $~~~~~~~~~~~~~~~~~~~~~~$[NC=85]:~P =(0.5195E-02,-.5135E-03)\\
 H1 CH+ CH-~~~~~~~~~~~[NC=86]:~G =(-.4047E-01,0.0000E+00)\\
 H1 CH+ W-~~~~~~~~~~~~[NC=87]:~G =(-.5024E-01,0.1030E-01)\\
 H1 photon photon~~~~~[NC=88]:~S =(-.6557E+01,0.2443E-01)\\
 $~~~~~~~~~~~~~~~~~~~~~~$[NC=89]:~P =(0.1385E-01,-.3423E-03)\\
 H1 glue glue~~~~(M=0)[NC=90]:~S =(0.1427E+01,0.1915E-17)\\
 $~~~~~~~~~~~~~~~~~~~~~~$[NC=91]:~P =(-.1291E-01,0.0000E+00)\\
 H1 photon photon(M=0)[NC=92]:~S =(-.4878E+01,0.5235E-17)\\
 $~~~~~~~~~~~~~~~~~~~~~~$[NC=93]:~P =(0.1728E-02,-.4811E-18)\\
 ---------------------------------------------------------\\
}
\item ${\tt IFLAG\_H(6)}=1$: The decay width and branching fractions of the 
lightest Higgs boson \\ {\tt
 ---------------------------------------------------------\\
 Neutral~Higgs~Boson~Decays~with~ISMN~=~~50~:~ISUSYN =~~50\\
 ---------------------------------------------------------\\
 DECAY~MODE~~~~[~IM]~~~WIDTH[GeV]~~BR[SM]~~~~~~BR[TOTAL]\\
 ---------------------------------------------------------\\
 H1 ->~e~~~~e~~[~~1]:~~0.3070E-10~~0.5766E-08~~0.5760E-08\\
 H1 ->~mu~~~mu~[~~2]:~~0.1393E-05~~0.2616E-03~~0.2613E-03\\
 H1 ->~tau~~tau[~~3]:~~0.3873E-03~~0.7274E-01~~0.7266E-01\\
 H1 ->~d~~~~d~~[~~4]:~~0.1686E-07~~0.3166E-05~~0.3163E-05\\
 H1 ->~s~~~~s~~[~~5]:~~0.6193E-05~~0.1163E-02~~0.1162E-02\\
 H1 ->~b~~~~b~~[~~6]:~~0.4163E-02~~0.7820E+00~~0.7812E+00\\
 H1 ->~u~~~~u~~[~~7]:~~0.2632E-08~~0.4944E-06~~0.4939E-06\\
 H1 ->~c~~~~c~~[~~8]:~~0.1124E-03~~0.2111E-01~~0.2109E-01\\
 H1 ->~t~~~~t~~[~~9]:~~0.0000E+00~~0.0000E+00~~0.0000E+00\\
 H1 ->~W~~~~W~~[~10]:~~0.4106E-03~~0.7711E-01~~0.7703E-01\\
 H1 ->~Z~~~~Z~~[~11]:~~0.3303E-04~~0.6203E-02~~0.6197E-02\\
 H1 ->~H1~~~Z~~[~12]:~~0.0000E+00~~0.0000E+00~~0.0000E+00\\
 H1 ->~H2~~~Z~~[~13]:~~0.0000E+00~~0.0000E+00~~0.0000E+00\\
 H1 ->~H1~~~H1~[~14]:~~0.0000E+00~~0.0000E+00~~0.0000E+00\\
 H1 ->~H1~~~H2~[~15]:~~0.0000E+00~~0.0000E+00~~0.0000E+00\\
 H1 ->~H2~~~H2~[~16]:~~0.0000E+00~~0.0000E+00~~0.0000E+00\\
 H1 ->~ph~~~ph~[~17]:~~0.9427E-05~~0.1771E-02~~0.1769E-02\\
 H1 ->~gl~~~gl~[~18]:~~0.2004E-03~~0.3763E-01~~0.3760E-01\\
 H1 TOTAL(SM)~~[~50]:~~0.5324E-02~~0.1000E+01~~0.9990E+00\\
 H1 ->~N1~~~N1~[~51]:~~0.5557E-05~~0.1044E-02~~0.1043E-02\\
 H1 ->~N1~~~N2~[~52]:~~0.0000E+00~~0.0000E+00~~0.0000E+00\\
 H1 ->~N1~~~N3~[~53]:~~0.0000E+00~~0.0000E+00~~0.0000E+00\\
 H1 ->~N1~~~N4~[~54]:~~0.0000E+00~~0.0000E+00~~0.0000E+00\\
 H1 ->~N2~~~N2~[~55]:~~0.0000E+00~~0.0000E+00~~0.0000E+00\\
 H1 ->~N2~~~N3~[~56]:~~0.0000E+00~~0.0000E+00~~0.0000E+00\\
 H1 ->~N2~~~N4~[~57]:~~0.0000E+00~~0.0000E+00~~0.0000E+00\\
 H1 ->~N3~~~N3~[~58]:~~0.0000E+00~~0.0000E+00~~0.0000E+00\\
 H1 ->~N3~~~N4~[~59]:~~0.0000E+00~~0.0000E+00~~0.0000E+00\\
 H1 ->~N4~~~N4~[~60]:~~0.0000E+00~~0.0000E+00~~0.0000E+00\\
 H1 ->~C1+~~C1-[~61]:~~0.0000E+00~~0.0000E+00~~0.0000E+00\\
 H1 ->~C1+~~C2-[~62]:~~0.0000E+00~~0.0000E+00~~0.0000E+00\\
 H1 ->~C2+~~C1-[~63]:~~0.0000E+00~~0.0000E+00~~0.0000E+00\\
 H1 ->~C2+~~C2-[~64]:~~0.0000E+00~~0.0000E+00~~0.0000E+00\\
 H1 ->~ST1*~ST1[~65]:~~0.0000E+00~~0.0000E+00~~0.0000E+00\\
 H1 ->~ST1*~ST2[~66]:~~0.0000E+00~~0.0000E+00~~0.0000E+00\\
 H1 ->~ST2*~ST1[~67]:~~0.0000E+00~~0.0000E+00~~0.0000E+00\\
 H1 ->~ST2*~ST2[~68]:~~0.0000E+00~~0.0000E+00~~0.0000E+00\\
 H1 ->~SB1*~SB1[~69]:~~0.0000E+00~~0.0000E+00~~0.0000E+00\\
 H1 ->~SB1*~SB2[~70]:~~0.0000E+00~~0.0000E+00~~0.0000E+00\\
 H1 ->~SB2*~SB1[~71]:~~0.0000E+00~~0.0000E+00~~0.0000E+00\\
 H1 ->~SB2*~SB2[~72]:~~0.0000E+00~~0.0000E+00~~0.0000E+00\\
 H1 ->STA1*STA1[~73]:~~0.0000E+00~~0.0000E+00~~0.0000E+00\\
 H1 ->STA1*STA2[~74]:~~0.0000E+00~~0.0000E+00~~0.0000E+00\\
 H1 ->STA2*STA1[~75]:~~0.0000E+00~~0.0000E+00~~0.0000E+00\\
 H1 ->STA2*STA2[~76]:~~0.0000E+00~~0.0000E+00~~0.0000E+00\\
 H1 ->SNU3*SNU3[~77]:~~0.0000E+00~~0.0000E+00~~0.0000E+00\\
 H1 TOTAL(SUSY)[100]:~~0.5557E-05~~0.1044E-02~~0.1043E-02\\
 H1 TOTAL~~~~~~[101]:~~0.5330E-02~~0.1001E+01~~0.1000E+01\\
  \\
 $*$~Note~:~WIDTH=GAMBRN(IM,1,1),~BR[SM]~~~=GAMBRN(IM,2,1) \\
 $~~~~~~~~~~~~~~~~~~~~~~~$and~~~~~~BR[TOTAL]=GAMBRN(IM,3,1) \\
 ---------------------------------------------------------\\
}
\end{itemize}
\end{appendix}

\end{document}